\documentclass[fleqn,usenatbib]{mnras}
\usepackage{newtxtext,newtxmath}
\usepackage[T1]{fontenc}
\usepackage{ae,aecompl}
\usepackage{graphicx}	% Including figure files
\usepackage{amsmath}	% Advanced maths commands
\usepackage{amssymb}	% Extra maths symbols

\newcommand{\jump}[1]{\ensuremath{[\![#1]\!]} }

\newcommand{\diff}{\mathrm{d}} %d in differentials
\newcommand{\dw}{\mathrm{(d)}}
\newcommand{\up}{\mathrm{(u)}}
\defcitealias{LG}{LG00}
\defcitealias{TY}{TY14}
\defcitealias{CritFluc}{NYY13}
\defcitealias{YY}{YY16}
\defcitealias{WHW02}{WHW02}

\title[Importance of progenitor asymmetry]{On the importance of progenitor asymmetry to shock revival in core-collapse supernovae}

\author[H. Nagakura et al.]{
Hiroki Nagakura,$^{1,2}$\thanks{E-mail: hirokin@princeton.edu}
Kazuya Takahashi$^{3}$
and Yu Yamamoto$^{4}$
\\
$^{1}$Department of Astrophysical Sciences, Princeton University, Princeton, NJ 08544\\
$^{2}$TAPIR, Walter Burke Institute for Theoretical Physics, Mailcode 150-17, California Institute of Technology, Pasadena, CA 91125, USA\\
$^{3}$Center for Gravitational Physics, Yukawa Institute for Theoretical Physics, Kyoto University, Kyoto, 606-8502, Japan\\
$^{4}$Advanced Research Institute for Science and Engineering, Waseda University, 3-4-1 Okubo, Shinjuku, 169-8555, Japan
}

\date{Accepted XXX. Received YYY; in original form ZZZ}

\pubyear{2015}

\begin{document}
\label{firstpage}
\pagerange{\pageref{firstpage}--\pageref{lastpage}}
\maketitle

\begin{abstract}
The progenitor stars of core-collapse supernovae (CCSNe) are asymmetrically fluctuating due to turbulent convections in the late stages of their lives. The progenitor asymmetry at the pre-supernova stage has recently caught the attention as a new ingredient to facilitate shock revival in the delayed neutrino-heating mechanism. In this paper, we investigate the importance of the progenitor asymmetries to shock revival with a semi-analytical approach. Free parameters were chosen such that the time evolution of shock radii and mass accretion rates are compatible with the results of detailed numerical simulations of CCSNe in spherical symmetry. We first estimate the amplitude of asymmetries required for the shock revival by the impulsive change of pre-shock flows in the context of neutrino heating mechanism, and then convert the amplitude to the corresponding amplitude in the pre-supernova phase by taking into account the growth of asymmetries during infall. We apply our model to various types of progenitors and find that the requisite amplitude of pre-supernova asymmetry is roughly three times larger than the prediction by current stellar evolution models unless other additional physical ingredients such as multi-dimensional fluid instabilities and turbulent convections in post-shock flows aid shock revival. We thus conclude that progenitor asymmetries can not trigger the shock revival by the impulsive way but rather play a supplementary role in reality.
\end{abstract}

%This suggests that progenitor asymmetries play only a supplementary role, rather than a primary role, for the shock revival in reality.
%We also find that the required fluctuations roughly increase with the mass shell. This means that we need to give an attention not only to the amplitude but also to the {\it locations} of progenitor asymmetries, which is a new insight into the link between stellar structures and the explosion mechanism.

\begin{keywords}
supernovae: general -- stars: evolution -- turbulence
\end{keywords}

%%%%%%%%%%%%%%%%%%%%%%%%%%%%%%%%%%%%%%%%%%%%%%%%%%

%%%%%%%%%%%%%%%%% BODY OF PAPER %%%%%%%%%%%%%%%%%%

\clearpage

\section{Introduction}\label{sec:intro}
The explosion mechanism of core-collapse supernovae (CCSNe) has been a long-standing problem despite decades of effort. The central issue in the theory of CCSNe is how a stagnated shock wave in a stellar core can overwhelm the ram pressure of accreting matter and be relaunched. The most promising energy supplier is supposed to be neutrinos that diffuse out from the central proto-neutron star (PNS) and transfer energy into the post-shock region \citep{JankaRev,Kotake12}. However, sophisticated numerical modelling of CCSNe revealed that the neutrino-heating process cannot revive the shock wave alone \citep{RJ00,Mezzacappa01,Liebendorfer05,Sumiyoshi,Suwa}, which means that there is still some missing physics in the explosion mechanism.

Based on the neutrino heating mechanism, various additional physical ingredients have been proposed to facilitate explosions. For instance, multi-dimensional (multi-D) fluid instabilities such as the standing accretion shock instability \citep[SASI;][]{Blondin03} and neutrino-driven convection kinetically push a shock wave outward and increase the efficiency of neutrino heating at the same time \citep[][and references therein]{FoglizzoRev,Mezzacappa15,MuellerRev}. Stellar rotation may be an important factor to foster shock expansion \citep{Nakamura14,Iwakami14b,Takiwaki16,2018ApJ...852...28S} while magnetic fields play more important roles for rapidly rotating progenitors \citep{Kotake04,Burrows07,Sawai14,2017MNRAS.469L..43O,Moesta,Sawai16}.
% The acoustic waves generated by g-mode oscillations of a PNS could be another energy source {\bf \citep{Burrows06,Shijun07,Harada17}.}
% (\citet{Burrows06,Shijun07,Harada17}, but see \citet{2008MNRAS.387L..64W} for counter arguments).
 These additional elements do not ensure a successful shock revival, however, except for the extreme ones such as very rapid rotation. This is indicated by the fact that state-of-the-art simulations by various groups show qualitatively different results \citep[][and references therein]{JankaRevNew}. It is hence indispensable to tackle a problem what ingredient is essential for CCSNe by improving numerical simulations while looking for other possible keys for shock revival that may have been missed.

More recently, turbulent fluctuations at the pre-supernova stage grab the spotlight as a new key for facilitating shock revival. Asymmetric fluctuations naturally inhere in the progenitors due to the development of violent convection in the burning Si/O shells \citep{Arnett94,Bazan,Asida,Meakin06,Meakin07,Arnett,Chat,Couch3,Muller16,Chat16,Jones}. Hence, generally speaking, they should be taken into account for all the progenitors of CCSNe. Indeed, it was recently found in numerical simulations that the upstream asymmetries manifestly increase the average shock radius and even lead an explosion for some progenitors \citep{Couch,MJ14,Couch2,Couch3,Burrows16,Muller17}.

Although it has become almost a consensus that progenitor asymmetries ease shock revival in the delayed neutrino heating mechanism, there are several issues which should be addressed. One of them is the uncertainty in the estimated amplitude of asymmetries. For the moment, the amplitude can be predicted only by the theory of stellar evolution due to the lack of observations. Even in the most advanced calculations of stellar evolution, however, various approximations and phenomenological treatments have been employed thus far. For instance, the matter profile is usually assumed to be spherically symmetric and multi-D effects such as mixing of elements are approximately handled by the mixing length theory \citep{Kippenhahn} although it has been pointed out that the standard mixing length theory misses the physics of convective boundary mixing \citep{Meakin07,Cristini}. Multi-d stellar evolution has been studied by hydrodynamical simulations for a short period of the final stage of evolution \citep{Bazan,Asida,2003ASPC..293..147K,Meakin06,Meakin07,Arnett,Chat,Couch3,Muller16,Jones,Muller17}. They found that thermodynamical quantities in Si/O shells before the onset of collapse would fluctuate asymmetrically roughly less than $10\%$ of their angle average. For instances, \citet{Bazan} found that $\lesssim 8 \%$ density perturbation at the edge of convective zone in their 2D hydrodynamical simulations. \citet{Asida} extended the simulation of \citet{Bazan} and they confirmed that the density fluctuation exists persistently in the convective boundary. \citet{Meakin06,Meakin07} carried out simulations of a late evolution of $23 M_{\odot}$ in 2D and 3D, and they found that 2D convective motions are exaggerated  than 3D by a factor of $\sim 8$. Such first-principles approaches to multi-D stellar evolution will rapidly mature as computational resources increase and give us the amplitude of asymmetries more accurately.

Another issue to be addressed is the dynamical role of the progenitor asymmetries in the post-bounce phase, which is not fully understood yet. There are at least two different ways they can impact the dynamics. The first one is the following direct way. Once asymmetric fluctuations hit the stalled shock surface, they break the force balance between the fluid upstream and downstream of the shock wave. This makes the transition from a quasi-steady state to a dynamical one. There exists a critical amplitude of fluctuations for which the shock wave eventually revives \citep[][hereafter \citetalias{CritFluc}]{CritFluc}, i.e., the progenitor asymmetry may be a key to directly triggering a shock revival if they have sufficiently large amplitudes. The chance is increased further by the fact that upstream asymmetries go through being amplified in supersonic accretion \citep[][hereafter \citetalias{TY}]{LG,TY}. If this is the case, the progenitor asymmetry could be more crucial for CCSNe than ever thought. Since the progenitor asymmetry and the amplification rate during infall would depend on the progenitor, it is necessary to systematically study many types of CCSN progenitors. However, CCSN simulations are, in general, computationally expensive and, in addition, uncertainties in stellar evolution models prevent us from developing physically correct initial conditions for these simulations. For these reasons, the possibility of this scenario has not been investigated in detail so far.

The other contribution from the progenitor asymmetries to shock revival is more complex and indirectly associated with the shock dynamics. Once the upstream asymmetric fluctuations pass through the shock wave, they couple with fluid instabilities and disturb post-shock accretion flows. As a result, the upstream fluctuations further intensify the neutrino heating and turbulent pressure and push the shock wave outward \citep{MJ14,2016ApJ...831...75T}
% \citep{Couch,2016ApJ...831...75T,Abd2}
 while the inherent fluid instabilities in the post-shock region such as SASI and neutrino-driven convections have already enhanced them. \citet{Abd2} showed that the total kinetic energy in the post-shock flows is amplified by a factor of $\sim 2$, which results in decreasing several percents of the critical neutrino luminosity. Since the post-shock flow is highly turbulent, it is difficult to quantify the effect of asyemmtrical flow on shock revival \citep{MM} as induced by infalling perturbations using detailed numerical simulations. Especially, numerical resolution may be a concern. For instances, \citet{HighRes} changes the grid resolution on their 3D simulations by factor 20 between the lowest and highest resolutions (the lowest resolution is similar to those used in other 3D simulations in \citet{2015ApJ...807L..31L,2015ApJ...801L..24M}). Although integral quantities such as the turbulent kinetic energy and average turbulent Mach number are not sensitive to resolutions, they found that the convections in the low simulations suffer from the so-called bottleneck effect which may facilitate the shock revival. (see e.g., Fig.2 in \citet{HighRes}). In addition, neutrino transport and the feedback to matter should be taken into account simultaneously and accurately, which makes simulations more computationally expensive and complicated. It is hence one of grand challenges in computational astrophysics. It is necessary to keep grappling with this problem by improving numerical schemes of CCSN simulations.

% since, according to \citet{HighRes}, even the most up-to-date 3D simulations are under-resolved.

\begin{figure*}
\includegraphics[width=\textwidth]{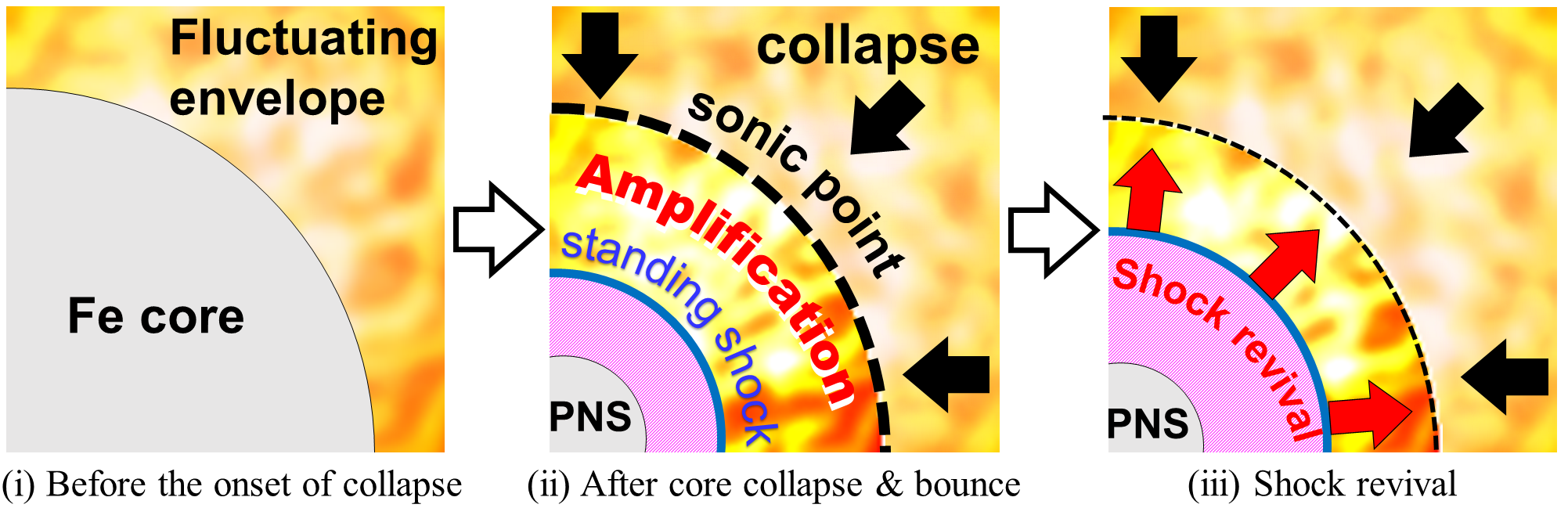}
\caption{A schematic picture of the evolutionary path to shock revival considered in this paper. The left panel displays the stellar structure right before the onset of core collapse. The envelope around an iron core is fluctuating due to turbulent convections with nuclear burning, which is shown as a mottled area in this picture. The middle panel shows the post-bounce phase. The PNS is enclosed with a stagnated shock wave while the outer envelope falls onto the shock wave. The fluctuation in the outer envelope is amplified during the accretion from the sonic point to the standing shock. The right panel shows the shock revival phase.}
\label{zukai}
\end{figure*}

In this paper, we examine the former scenario in which progenitor asymmetries directly and crucially aid shock revival based on the neutrino heating mechanism. Although we faced the practical and theoretical obstacles mentioned above, we overcome them by using a novel approach. The fundamental elements in our method consist of three semi-analytical approaches in \citet{BG93}; \citetalias{CritFluc}; \citetalias{TY}. Our method is less expensive than numerical simulations and, more importantly, free parameters in our model are chosen so as to reproduce detailed numerical simulations. After establishing reliability, we proceed to apply our models to various types of progenitors. To expedite the understanding of progenitor dependence, we employ parametrically generalized progenitor models in \citetalias{YY} in addition to some representative CCSNe progenitors computed by realistic stellar evolution calculations.

%rather than realistic models provided by 
% To see the progenitor dependence, 
%They are a series of a 
% various types of progenitors.

 Another important point of our study is that we reverse the standard approach to the problem. We first estimate the amplitude of progenitor asymmetry necessary for shock revival and then compare it with a canonical amplitude in calculations of stellar evolution. Our study does not need the information on the accurate amplitude in advance but rather give constraints on them. Note also that we compute the necessary progenitor asymmetries as a function of mass coordinate. It yields information about the locations where asymmetries are more important for shock revival.
% which is a new insight into the link between the stellar structures at the pre-supernova stage and the explosion mechanism.

This paper is organized as follows. To facilitate the readers' understanding, we start out with an overview of our study in Sec.~\ref{sec.overview}. We describe our method in Sec.~\ref{sec.method} and progenitor models in Sec~\ref{sec.prog}. The main results are shown in Sec.~\ref{sec.results} while the limitations of the current study are examined in Sec.~\ref{sec.limitation}. Finally, Sec.~\ref{sec.sum} concludes the paper with summary and discussion.

\section{Overview of this study}\label{sec.overview}
In this section, we sketch the evolutionary path to shock revival considered in this paper.
% in which the shock revival is driven by the neutrino heating mechanism aided primarily by progenitor asymmetries.
As described in the left panel of Fig.~\ref{zukai}, all hydrodynamical quantities outside an iron core are asymmetrically fluctuating due to turbulent convections before the onset of collapse. These fluctuating envelopes fall by the rarefaction wave that is generated by collapse of the iron core. As a result of collapse, a PNS forms at the center and, at the same time, a shock wave is generated and propagates through the accreting matter. Since the expansion of shock wave is interrupted by the neutrino cooling and photodissociation of heavy nuclei, the system settles in a quasi-steady state at $\sim 100$ms after the bounce. In this phase, the position of shock wave is mainly determined by the balance between the ram pressure of the upstream flow and the thermal pressure of the post-shock flow aided by neutrino heating \citep[See e.g.,][]{BG93,Ohnishi,Yamasaki06}. Meanwhile, in the pre-shock flow, the asymmetric fluctuations of the outer envelopes increase during infall and then eventually be swallowed in the shock wave as depicted in the middle panel of Fig.~\ref{zukai}. If the fluctuations are significantly large, they potentially trigger a shock revival as shown in the right panel of the same figure.

What we estimate in this study is how large pressure fluctuations in Si/O shells at pre-supernova stage are required for shock revival. We denote the minimally required amplitude by $f_\mathrm{crit}^\mathrm{Si/O}$ (the precise definition is given later).
% Finally we compare $f_\mathrm{crit}^\mathrm{Si/O}$ with corresponding values in calculations of stellar evolution and then diagnose the importance of progenitor asymmetry.
%{\it before the onset of collapse}
% As explained later, the critical value can be obtained once the shock evolution is given. For this purpose, we use results of 1D numerical simulations with the full-Boltzmann neutrino transport (Nagakura et al.). In addition, we take into account the diversity of the evolution of supernova cores due to progenitor difference and numerical methods by modeling the shock evolution in an analytic way as briefly described below.
%The calculation of $f_\mathrm{crit}^\mathrm{Si/O}$ consists of two steps, which are briefly described below.
We compute $f_\mathrm{crit}^\mathrm{Si/O}$ by multiple steps. At first we apply a semi-analytical model for the post-bounce phase of CCSNe, which is built based on a quasi-steady approximation with light-bulb neutrino transport \citep{BG93, Ohnishi, Yamasaki06}. Given three characteristic quantities; mass accretion rate ($\dot{M}$), mass of the PNS ($M_\mathrm{PNS}$) and neutrino luminosity ($L_\nu$), we apply the quasi-steady model to obtain the time evolution of shock wave and mass of PNS, both of which are the necessary quantities to measure the critical fluctuation for shock revival ($f_\mathrm{crit}$, see below). The characteristics of accretion flows are encompassed in $\dot{M}$, $M_\mathrm{PNS}$ and $L_\nu$, which are computed from the density profile of progenitors at the pre-supernova stage. Note also that some free parameters in our quasi-steady model are calibrated with the result of more detailed numerical simulations.

% Note that there are free parameters in the quasi-steady model (e.g., the conversion efficiency of accretion energy to neutrinos). They are chosen so as to reproduce the results of realistic numerical simulations.

%The post-shock flow is characterised by three control parameters, which are the neutrino luminosity ($L_\nu$), the mass accretion rate ($\dot{M}$) and the mass of the PNS ($M_\mathrm{PNS}$). They are computed from the density profile of progenitors at the pre-supernova stage according to an elaborate model explained in the next section and Appendix~\ref{LMmodel}.

 For a given quasi-steady evolution obtained in the previous step, we then proceed the next step to estimate the minimal fluctuation required for shock revival ($f_\mathrm{crit}$) at the shock point by applying the result of \citetalias{CritFluc}. Importantly, $f_\mathrm{crit}$ is different from $f_\mathrm{crit}^\mathrm{Si/O}$, since the asymmetries are amplified during infall, i.e., $f_\mathrm{crit}$ is generally larger than $f_\mathrm{crit}^\mathrm{Si/O}$. We employ the scaling law in \citetalias{TY} to compute the amplification rate and then convert $f_\mathrm{crit}$ to $f_\mathrm{crit}^\mathrm{Si/O}$. We finally assess the feasibility of $f_\mathrm{crit}^\mathrm{Si/O}$ by comparing the prediction from current stellar models, and assess the importance of progenitor asymmetry for shock revival.

% compare the obtained $f_\mathrm{crit}^\mathrm{Si/O}$ with the 
% the importance of progenitor asymmetry to shock revival. In the following section, we describe the details of each step in order.

\section{Method}\label{sec.method}

\subsection{Step 1: Shock evolution in the supernova core}\label{subsec.quasisteady}
%\subsection{Step 1: Quasi-steady evolution of the supernova core}\label{subsec.quasisteady}
As to be cleared later, the critical amplitude of accretion asymmetries for shock revival is related to the shock radius and the mass of PNS. Although the numerical simulation is a straightforward way to obtain them, it usually takes a huge computational cost because of the complexity of CCSN physics. We hence employ a semi-analytical approach based on a quasi-steady model with a light-bulb neutrino transport. Below we describe the method in detail.

% to compute the time evolution of shock wave
% We, however, simulated the shock evolutions for two types of canonical supernova progenitors with the solar metallicity of \citet[][hereafter \citetalias{WHW02}]{WHW02}, s15 and s27, by using a state-of-the-art code employing full-Boltzmann neutrino transport \citep{Nagakura18b} with an EoS based on realistic nuclear forces \citep{Togashi14,Togashi17,Furusawa17}.

%In order to catch the variety of shock evolution for different progenitors, we propose a model that produces shock evolution in the supernova core for a given progenitor star. The model is applied to additional six types of progenitors mentioned later as well as s15 and s27 progenitors to produce various shock evolutions after a calibration of parameters with our 1D full-Boltzmann simulations. The readers can confirm the reproducibility of shock evolutions in Fig.~\ref{fig.comp}.
%After the shock is stalled, the accretion flow onto the PNS is roughly in a quasi-steady state and its evolution can be well approximated as a series of steady states.

%supported by the fact that the evolution of the supernova core after shock stall is, indeed, 

The quasi-steady model employed in this paper was originally developed by \citet{BG93} and then being extended by \citet{Yamasaki06}. Mathematically speaking, this model is built on a solution of spherically symmetric time-independent fluid equations. There are three characteristic quantities to define the structure of shocked accretion flows, which are the mass accretion rate ($\dot{M}$), the mass of the PNS ($M_\mathrm{PNS}$) and the neutrino luminosity ($L_\nu$). The quasi-steady approximation is justified by the fact that these characteristic quantities evolve sufficiently slowly than the dynamical time scale of post-shock flows.
%The time scale of the evolution of these quantities is much longer than the dynamical time scale of the system
% well approximated as a series of steady states, since accretion flow onto the PNS is roughly in a quasi-steady state.
%\citet{BG93} developed a phenomenological model to describe the steady accretion flow as a solution of spherically symmetric time-independent fluid equations where constant $L_\nu$, $\dot{M}$ and $M_\mathrm{PNS}$ are assumed.
 In spite of the simplicity, this approach is frequently used in the literature to study the property of accretion flows qualitatively. In addition to this, the results are also used to measure the closeness for shock revival in numerical simulations (see e.g., \citet{MB08,Hanke12}).

%, 
% Indeed, complex dynamics in numerical simulations becomes more understandable by using the model \citep{MB08,Hanke12}. In this study, we also use an essentially same model while we develop further extensions by including progenitor characteristics into three control parameters ($L_\nu$, $\dot{M}$ and $M_\mathrm{PNS}$).

Assuming the accretion flow is steady and spherical symmetric, the basic equations of fluid with neutrino interactions are written as
\begin{align}
\label{eq1}
&\dot{M} = 4\pi r^2 \rho v = \mathrm{const.}, \\
\label{eq2}
&v\frac{\diff v}{\diff r} = -\frac{1}{\rho} \frac{\diff p}{\diff r} -\frac{GM_\mathrm{PNS}}{r^2}, \\
\label{eq3}
&v\frac{\diff \varepsilon}{\diff r} +pv\frac{\diff }{\diff r}\left( \frac{1}{\rho} \right) = q(L_\nu), \\
\label{eqLast}
&\rho v\frac{\diff Y _\mathrm{e}}{\diff r} = \lambda(L_\nu),
\end{align}
where $r$, $\rho$, $v$, $p$, $G$, $\varepsilon$ and $Y_\mathrm{e}$ are the radius, baryon mass density, radial velocity, pressure, gravitational constant, specific internal energy and electron fraction, respectively. For gravity, we take a monopole approximation of PNS and also ignore the self-gravity in accretion flows. We ignore the momentum exchange between neutrinos and matter just for simplicity. $q$ and $\lambda$ are the neutrino heating and deleptonization rates, respectively. They are function of the neutrino luminosity $L_\nu$, which is introduced later. We employ Shen equation of state (EoS) \citep{Shen} which is one of the frequently used nuclear EoS\footnote{Note that Shen EoS does not satisfy the current observational constraint of mass-radius relation of neutron star (see e.g., \citep{2013ApJ...774...17S}). However, this study focuses only on the subnuclear density regime ($\lesssim 10^{11}{\rm g/cm}^3$) in which all thermodynamical quantities and chemical abundances are almost the same as other realistic supernova EoSs.}. The rest of other variables, $\dot{M}$, $M_\mathrm{PNS}$ and $L_\nu$ in the above equations define the characteristics of accretion flows and they are evaluated before solving Eqs.~(\ref{eq1})-(\ref{eqLast}). As shown below, these parameters can be written as a function of mass-shell ($M_r$) of progenitors. We also relate the post-bounce time ($t_{\rm pb}$) to $M_r$ through $t_\mathrm{infall}$ (see Eq.~(\ref{deftpb})). In other words, $t_{\rm pb}$ can be measured by what mass shell $M_r$ falls on to the shock surface.

 Given $\dot{M}$, $M_\mathrm{PNS}$ and $L_\nu$, Eqs.~(\ref{eq1})-(\ref{eqLast}) are solved until the solution satisfies boundary conditions. The outer boundary condition is given at the shock surface by Rankine-Hugoniot condition. The pre-shock accretion flow is approximated as an adiabatic flow with a free-fall velocity. The inner boundary condition is given at the neutrino sphere ($r_{\nu}$) which is defined at the position with $\rho = 10^{11}$~g~cm$^{-3}$ in this study. On the other hand, $r_{\nu}$ is also related with neutrino luminosity (see Eq.~(\ref{eqLbound})) which indicates that $r_{\nu}$ should satisfy both conditions simultaneously. We iteratively solve Eqs.~(\ref{eq1})-(\ref{eqLast}) in the region surrounded by the neutrino sphere and the shock radius until $r_{\nu}$ satisfies both conditions.
% This type of problem corresponds to an eigen-value problem on cardinally differential equations.
% Below, we give a formal definitions and expressions for important quantities.

% functions of $M_r$ for three characteristic variables.
% (the detail of these formulations are described below)

% from a given shock radius to the neutrino sphere. The shock radius, $r_\mathrm{sh}$, is determined so as 
%$t_\mathrm{infall}$, which is the infall timescale for mass shells at the pre-supernova stage (See Eq. *** for the definition). 
%The time evolution of the system can be obtained by changing these parameters.

Below we describe how $\dot{M}$, $M_\mathrm{PNS}$ and $L_\nu$ can be determined from the progenitor structure. The mass accretion rate $\dot{M}$ at the shock surface can be approximately computed by the radial distribution of density profile before the onset of collapse. Following \citet{Nagakura,2015ApJ...806..145W,Suwa}, we compute it as
\begin{equation}
\label{Mdot}
\dot{M} =\frac{\diff M_r}{\diff r}\left(\frac{\diff t_\mathrm{infall}}{\diff r}\right)^{-1},
\end{equation}
where the infall timescale, $t_\mathrm{infall}$, is given by
\begin{equation}
\label{tff}
t_\mathrm{infall} = \alpha \sqrt{\frac{r^3}{GM_r}}.
\end{equation}
Here, $r=r(M_r)$ is the radial coordinate of the mass shell at the onset of collapse. $\alpha$ is a deviation factor from the free-fall timescale, which is calibrated from numerical simulations\footnote{We introduce the deviation factor $\alpha$ since the stellar pressure contributes to suppress the infall velocity.}. By substituting Eq.~(\ref{tff}) and $ \diff M_r/\diff r= 4\pi r^2\rho$ into Eq.~(\ref{Mdot}), the mass accretion rate of the mass shell at $M_r$ is given by
\begin{equation}
\label{Mdot2}
\dot{M} = \frac{1}{\alpha^2}\frac{8\pi GM_r^2 t_\mathrm{infall}\rho}{3M_r -4\pi r^3\rho},
\end{equation}
where $\rho(M_r)$ denotes the matter density at the corresponding mass shell at the onset of collapse.

$M_\mathrm{PNS}$ can be related with $M_r$ by taking a following approximation. Since the mass of PNS dominates the total mass inside the shock wave, we ignore the mass between shock wave and PNS surface just for simplicity. Thus we obtain the relation
\begin{equation}
\label{Mpns}
M_\mathrm{PNS} =  M_r.
\end{equation}
In other words, we treat $M_\mathrm{PNS}$ by the mass shell that hits the shock wave at that time.
%This approximation is valid in $t_{\rm{b}} \gtrsim 100$ms since most of accretion flows in the post-shock region has been already accreted onto the PNS.

Before describing how we determine $L_\nu$ as a function of $M_r$, we connect $t_{\rm pb}$ with $M_r$ through $t_\mathrm{infall}$. The connection is made with based on the idea which $t_{\rm pb}$ can be measured in terms of the time when the corresponding mass shell passes through the shock wave. Since the latter time relates with $t_\mathrm{infall}$, we can write $t_{\rm pb}$ as
\begin{equation}
\label{deftpb}
t_{\rm pb} =  t_\mathrm{infall}(M_r) - t_\mathrm{infall}(M_r^{\rm ref}),
\end{equation}
where $M_r^{\rm ref}$ denotes the reference mass shell to measure $t_{\rm pb}$. It is set as $M_r^{\rm ref}=1.42 M_\odot$, which reproduces the result of numerical simulations in our model.
% Owing to Eq.~(\ref{deftpb}), we can convert arbitral quantities as a function of $M_r$ to $t_{\rm pb}$.

% which is one of free parameters to be calibrated by numerical simulations.

The last characteristic quantity, $L_\nu$, can be determined as follows. At first, we divide the neutrino luminosity into two components: accretion and diffuse parts, i.e., we write $L_\nu$ as
\begin{equation}
\label{Lnu}
L_\nu = L_{\nu,\mathrm{acc}} + L_{\nu,\mathrm{diff}}
\end{equation}
 The former, in general, overwhelms the latter in early time of the post-bounce phase while it is reversed later \citep{Summa,Suwa}. Since the time evolutions of these two components obey different physics, they are separately treated in our model. The accretion luminosity can be written as
\begin{equation}
\label{Lacc}
%L_{\nu,\mathrm{acc}} = \eta \frac{GM_r}{R_\mathrm{acc}} \min(\dot{M},1~M_\odot~\mathrm{s}^{-1}).
L_{\nu,\mathrm{acc}} = \eta \frac{GM_r}{R_\mathrm{acc}}\dot{M}.
\end{equation}
In the above equations, $\eta$ denotes the conversion efficiency from the accretion kinetic energy to the neutrino energy. The radius, $R_\mathrm{acc}$, represents a typical location where some dissipative processes occurs and then convert the kinetic energy to thermal energy with neutrino emissions. We set $R _\mathrm{acc}= 50$~km, which is assumed to be a typical radius of the PNS. Note that $\eta$ is one of free parameters in our model and being calibrated by comparing to numerical simulations (see Appendix~\ref{sec.calb}). In this sense, the uncertainty of $R_\mathrm{acc}$ to compute $L_{\nu,\mathrm{acc}}$ is also included in $\eta$. By substituting Eq.(\ref{Mdot2}) into Eq.~(\ref{Lacc}), we obtain $L_{\nu,\mathrm{acc}}$ as a function of $M_r$.

%In this sense, the uncertainty of $R_\mathrm{acc}$ is also included in $\eta$.

% The min function is introduce to limit the accretion luminosity for large $\dot{M} > 1 M_\odot$~s${}^{-1}$ expected in early phase just after the bounce when the shock radius is yet to increase: Without the artificial suppression of the accretion luminosity, our model turns to overestimate the shock radius for given $\dot{M}$ in the early phase. The phase shortly after the bounce would be, however, in a dynamical state rather than in a steady state and may not be treated with the following quasi-steady evolution. The suppression of $L_{\nu,\mathrm{acc}}$ is thus an ad-hoc prescription to roughly trace the shock trajectory even in the early phase, when the steady-state approximation would break down.

% rather than post-shock accretion and PNS structures.

For the diffuse component, on the other hand, it is difficult to be modelled prior to solving Eqs.~(\ref{eq1})-(\ref{eqLast}), since it is not determined only by progenitor structure. Instead, the time evolution of the diffusion component depends weakly on progenitors (see e.g., \citet{Summa}). It is attributed to the fact that the PNS structure is almost the same among all CCSN progenitors. Indeed, the neutrino luminosity decreases linearly in time in the late post-bounce phase and is not sensitive to progenitors as shown in detailed numerical simulations by \citet{Summa} (See Fig.~3 in their paper). For this reason, we approximately treat the time evolution of the neutrino diffuse component ($L_{\nu,\mathrm{diff}}$) as
\begin{equation}
\label{Ldiff}
L_{\nu,\mathrm{diff}}(t_{\rm pb}) = \dot{L} \hspace{0.5mm} t_{\rm pb} + L_{\nu,\mathrm{diff}}^\mathrm{ref},
\end{equation} 
where $\dot{L} (< 0)$ and $L_{\nu,\mathrm{diff}}^\mathrm{ref}$ denote a decline rate and a constant with respect to the time, both of which are model parameters to be calibrated from numerical simulations. Note that Eq.~(\ref{Ldiff}) appears at first glance to be no progenitor dependence. It is not true since the progenitor dependence is encompassed in $t_{\rm pb}$ (for instance, the post-bounce time of less compact Si/O shells evolves slowly with increasing $M_r$). By combining Eqs.(\ref{Lnu})-(\ref{Ldiff}), we finally obtain $L_\nu$ as a function of $M_r$.

As mentioned above, several free parameters should be calibrated in our model. We summarize the calibration in Appendix~\ref{sec.calb}. As a reference, we utilize the results of one of the most-recent CCSNe simulations in spherical symmetry \citep{Nagakura18b}. We confirm that our improved quasi-steady model gives us resonably consistent results with numerical simulations. We also check their parameter dependence in Appendix~\ref{appModel}.

%with a sufficient accuracy for the purpose of this paper

% within the possible errors of our calibration, and confirm that the uncertainties do not change our main conclusions in this paper.

% The light-bulb neutrino transport usually neglect heavy-leptonic neutrinos and only consider the chaged-current neutrino absorptions for electron-type neutrinos ($\nu_e$) and their anti-particles ($\bar{\nu}_e$) (see \citet{Ohnishi}). In addition to this, we further assume that luminosities and average energies between 

% both luminosities and average energies are insensitive to neutrino species

% for the former one, we take the light-bulb approximation \citep{Ohnishi, Scheck}, where the neutrino distribution function is simply determined by the distance from the effective neutrino sphere, instead of solving the neutrino transport.

Finally, we explain how to evaluate $q$ and $\lambda$ in Eqs.~(\ref{eq3}) and (\ref{eqLast}) which correspond to energy and lepton exchange between neutrinos and matter. For the dynamics of neutrinos, we use the light-bulb approximation instead of solving detailed neutrino transport in CCSNe. The essence of this approximation is that thermal neutrinos freely propagate outside of the neutrino sphere. Owing to the optically thin approximation, the distribution function of neutrinos ($f$) can be written in terms of thermal spectrum and geometrical factor for their angular distributions (see Eq.(18) in \citet{Ohnishi}\footnote{There is a typo in Eq.(18) of \citet{Ohnishi}. $2 \pi$ should be replaced to $2$.}). $f$ is directly used to evaluate $q$ and $\lambda$. In the canonical light-bulb approximation, only two (+ their inverse) weak processes are taken into account, which are the electron capture by free proton and the positron capture by free neutron \citep{Bruenn}. Their expressions are adopted from Eqs. (16) and (17) in \citet{Ohnishi}.

 Neutrino luminosities ($L_\nu$) and their temperatures ($T_{{\nu}_\mathrm{e}}$ and $T_{\bar{{\nu}}_\mathrm{e}}$) can be arbitrarily given in the light-bulb approximation. We apply Eq.~(\ref{Lnu}) to determine $L_\nu$. On the other hand, we set temperature of electron-type neutrinos ($\nu_e$) and their antiparticles ($\bar{\nu}_e$) as $T_{{\nu}_\mathrm{e}} = T_{\bar{{\nu}}_\mathrm{e}} = 4.5$ MeV just for simplicity. In the light-bulb approximation, neutrino luminosity can be written in terms of the radius of neutrino sphere ($r_{\nu}$) and neutrino temperature ($T_{\nu}$) as
\begin{equation}
L_{\nu(i)} = \frac{7}{16}4\pi r_{\nu(i)}^2 \sigma T_{\nu(i)}^4 , \label{eqLbound}
\end{equation}
where $\sigma $ and $i$ denote the Stefan-Boltzmann constant and the index of neutrino species ($i = \nu_e$ or $\bar{\nu}_e$), respectively. We further assume that both neutrino luminosities is the same between $\nu_e$ and $\bar{\nu}_e$, i.e., $L_{\nu_e} = L_{\bar{\nu}_e} = L_{\nu}$, and we also do not distinguish their neutrino spheres ($r_{\nu_e} = r_{\bar{\nu}_e}$).

Although these treatments of neutrino transport and feedback to matter are quite simplified, we compensate the weakness by calibrating free parameters. Indeed, our quasi-steady model succeeds to reproduce the realistic time evolution of shock radius and mass of PNS as shown in Appendix~\ref{sec.calb}. Note that our treatment of neutrino transport is not accurate enough to predict realistic neutrino signals from CCSNe, which is, however, out of the scope of this paper.

% by using quasi-steady approximation with light-bulb neutrino transport.
Here is summary of Step 1. Prior to solving basic equations for shocked-accretion flows (Eqs.~(\ref{eq1})-(\ref{eqLast})), we relate three characteristic quantities ($\dot{M}$, $M_\mathrm{PNS}$ and $L_\nu$) to $M_r$ by connecting them with the density profile of each progenitor. Given three characteristic quantities, we iteratively solve Eqs.~(\ref{eq1})-(\ref{eqLast}) until $r_{\nu}$ satisfies two conditions; $\rho(r_{\nu})=10^{11}$g/cm$^3$ and Eq.~(\ref{eqLbound}). The solution gives the shock radius ($r_\mathrm{sh}$). This means that we can obtain $r_\mathrm{sh}$ as a function of $M_r$ (or $t_{\rm pb}$). This is the final outcome of Step 1 and we use $r_\mathrm{sh}(M_r)$ and also $M_\mathrm{PNS}(M_r)$ for the next step.

Finally we define the mass shells which we consider in this paper. We focus on the mass shell in $1.5\lesssim M_r/M_\odot \lesssim 1.8$ which are swalled onto the shock wave at $0.1{\rm s} \lesssim t_{\rm pb}\lesssim 0.5$s in the case of s15 progenitor, for example. We exclude the phase in $M_r/M_\odot \lesssim 1.5$, i.e., very early phase of the post-bounce, since the shocked accretion flow does not settle to the quasi-steady state, which means that our model is not applicable. We also find that our quasi-steady model deviates from the result of numerical simulations when $\dot{M}$ is larger than $1~M_\odot~\mathrm{s}^{-1}$. The main reason of this deviation is that the accretion component of neutrino luminosity in our treatment becomes much larger than the reality. This problem can be overcome by recalibrating free parameters in the early post-bounce phase. However, we choose another away to avoid increasing the further complexity in our model. We take an ad hoc prescription in which we set the upper limit of the accretion component of neutrino luminosity. The upper limit is set as $\dot{M} =1~M_\odot~\mathrm{s}^{-1}$ in Eq.~(\ref{Lacc}). We also exclude the range $M_r/M_\odot \gtrsim 1.8$ since we again need to readjust the free parameters in our model. Albeit these caveats in our model, our model is still capable of covering the most important phase of CCSNe. Indeed, we expect to have a shock revival $t_{\rm pb}\lesssim 0.5$s for most of progenitors otherwise explosion energy would result in less than $10^{51}$erg \citep{Yamamoto13}.

\subsection{Step 2: Required fluctuations for shock revival at the shock surface}
The second step is more related to progenitor asymmetries. According to \citetalias{CritFluc}, there is a critical amplitude of fluctuations to revive a stalled shock wave (see Fig.~4 in their paper), which can be approximately given by
\begin{equation}\label{fcrit_in_body}
f_\mathrm{crit}^{\dw} \equiv \left. \frac{\delta p}{p} \right|_\mathrm{crit}^{\dw} \sim 0.8 \left( \frac{M_\mathrm{PNS}}{1.4\ \mathrm{M_\odot}} \right) \left[ 1 -\left( \frac{r_\mathrm{sh}}{10^8 \mathrm{\ cm}}\right)\right],
\end{equation} 
where $\delta p$ denotes the pressure fluctuation behind the shock wave. The superscript $\dw$ is assigned to denote clearly the side of downstream of shock wave. Note that $f_\mathrm{crit}^{\dw}$ increases for larger $M_\mathrm{PNS}$ or smaller $r_\mathrm{sh}$ because shock revival becomes more difficult under stronger gravitational field. By combining Eq.~(\ref{fcrit_in_body}) with the result of the previous step, $f_\mathrm{crit}^{\dw}$ can be also labeled on each $M_r$. Note that the critical amplitude given by Eq.~(\ref{fcrit_in_body}) contains errors which would be within $20 \%$ as checked by a 2D simulation in \citetalias{CritFluc}. Note that we will compare our criterion to others in Sec.~\ref{efffcrit}.

% We assume that a shock wave starts to explode once the magnitude of pressure fluctuation in the post-shock region exceeds the criterion $f_\mathrm{crit}^{\dw}$ by the upstream perturbation carried into the shock wave.

As shown in Eq.~(\ref{fcrit_in_body}), the critical fluctuation ($f_\mathrm{crit}^{\dw}$) is measured at the post-shock region. On the other hand, we currently consider the fluctuations driven by progenitor asymmetry, i.e., the asymmetric fluctuations are brought by the pre-shock accretion. Thus, we need to consider the conversion factor from the pre-shock fluctuation to the post-shock. In the following, we make a connection between pre-shock and post-shock fluctuations by using the linearized equation for the momentum conservation.

The linearized equation with respect to perturbed quantities for the momentum conservation between pre- and post-shock wave can be written as
	\begin{eqnarray}\label{eq.p}
	\left.\frac{\delta p}{p}\right|^\dw 
       &=&
         f_{V_\mathrm{sh}}\frac{\delta V_\mathrm{sh}}{v^\up}
       + f_{r_\mathrm{sh}}\frac{\delta r_\mathrm{sh}}{r_\mathrm{sh}} 
       + f_\rho \left.\frac{\delta \rho}{\rho}\right|^\up \nonumber \\
    && + f_v\left.\frac{\delta v}{v}\right|^\up
       + f_p\left.\frac{\delta p}{p}\right|^\up
       + f_{Y_e}\left.\frac{\delta Y_e}{Y_e}\right|^\up ,
	\end{eqnarray}
	where
	\begin{eqnarray}
	f_\rho &=& \frac{jv_1^\dw}{|A|p^\dw}\left\{\frac{1}{2}(\beta -1)^2 -\left[\frac{\varepsilon}{v^2}+(\beta-1)\frac{p}{jv}+(2-\beta)\frac{\rho}{v^2}\frac{\partial \varepsilon}{\partial \rho}\right]^\dw \right. \nonumber \\
	&&  \left.+ \beta^2 \left[\frac{\varepsilon}{v^2}+\frac{\rho}{v^2}\frac{\partial \varepsilon}{\partial \rho}\right]^\up \right\}, \label{eq:frho} \\
	f_v&=&\frac{jv_1^\dw}{|A|p^\dw}\left\{\frac{1}{2}(3\beta-1)(\beta-1) -\left[\frac{\varepsilon}{v^2}+(2\beta-1)\frac{p}{jv}\right.\right. \nonumber \\
	&& \left.\left. -2(\beta-1)\frac{\rho}{v^2}\frac{\partial \varepsilon}{\partial \rho}\right]^\dw + \beta^2 \left[\frac{\varepsilon}{v^2}+\frac{p}{jv}\right]^\up \right\}, \label{eq:frhv} \\
	f_p&=&\frac{p^\up}{|A|p^\dw}\left[\beta -1 +\left(\frac{\rho}{v^2}\frac{\partial \varepsilon}{\partial \rho}-\frac{p}{jv}\right)^\dw + \beta \rho ^\up \left.\frac{\partial \varepsilon }{\partial p}\right|^\up \right], \label{eq:fp}\\
	f_{Y_e}&=&-\frac{j\beta Y_e^\up}{|A|p^\dw v^{\up}}\jump{\frac{\partial \varepsilon}{\partial Y_e}}, \label{eq:fYe} \\
	f_{V_\mathrm{sh}} &=& -f_v \label{eq:fVsh}
	\\
	f_{r_\mathrm{sh}}&=&\frac{jv^\dw}{|A|p^\dw}\left[ \beta(\beta-1)\left(\frac{GM}{r_\mathrm{sh}v^{2\up}}-\frac{2}{\beta}\right)\left(-1+\frac{\rho}{v^2}\frac{\partial \varepsilon}{\partial \rho}-\frac{p}{jv}\right)^\dw \right. \nonumber \\
       &&  \left. +\frac{\beta^2}{v^{2\up}}\jump{\frac{2(E+p)}{\rho}-\frac{r_\mathrm{sh}q}{v}}-\frac{r_\mathrm{sh}m_B\jump{\lambda}}{jv^\dw}\left.\frac{\partial\varepsilon}{\partial Y_e}\right|^\dw \right] \label{eq:frsh}.
	\end{eqnarray}
Here, $\dw$ and $\up$ are superscripts to distinguish downstream and upstream quantities. $\delta V_\mathrm{sh}(=\diff r_\mathrm{sh}/\diff t)$, $j(=\rho^\up v^\up = \rho ^\dw v^\dw)$ and $\beta(=\rho ^\dw/\rho ^\up)$ denote the shock velocity, the unperturbed mass flux and the compression ratio of the unperturbed flow, respectively. The symbol $\jump{}$ is defined as $\jump{X}~=~X^\dw~-~X^\up$. $|A|$ is related with the determinant of the matrix for the coefficients of Rankine-Hugoniot relations for the perturbed flow, which is given by unpertubed quantities as
\begin{equation}
  |A|=\left(\rho \frac{\partial \varepsilon}{\partial p} - \frac{p}{jv_r} + \frac{\rho}{v_r^2}\frac{\partial \varepsilon}{\partial \rho}\right)^\dw.
\end{equation}
 Note that all coefficients denoted as $f_{*}$ in Eq.~(\ref{eq:frho})-(\ref{eq:frsh}) are determined by unperturbed quantities, and we find $|f_\rho|,\ |f_v| \gg |f_p|,\ |f_{Y_e}|$. This fact indicates that the asymmetric ram-pressure predominantly affect fluctuations in the post-shock flows. In the following analysis, we assume $\delta V_\mathrm{sh} = \delta r_\mathrm{sh} = 0$ just for simplicity\footnote{Strictly speaking, we should not set them as a priori but rather calculate them by solving Riemann problems as demonstrated in \citetalias{CritFluc}. It should be noted, however, that full perturbed quantities at the pre-shock (which are required to solve Riemann problem) can not be treated appropriately in this study since we employ the scaling law in \citetalias{TY} for the growth of perturbation during infall. The scaling law provides us only the representative magnitude of fluctuations. See the body of this paper for more details.}.

% to estimate the amplification of asymmetry during infall (see in Sec.~\ref{subsec.fluctu} in more details).

Since the upstream perturbations are in the same order for pressure, density and radial velocity (\citetalias{TY}), we introduce $x$ as the representative amplitude of perturbation\footnote{We exclude the term of $Y_e$ perturbation in Eq.~(\ref{eq:defx}). \citetalias{TY} does not estimate the order of $Y_e$ perturbation since they did not take into account weak interactions during infall. It should be noted however that $\delta Y_e$ would be smaller than perturbations of other quantities since the unperturbed Si/O shells go through less deleptonization during infall.},
\begin{equation}
|x|\equiv
\left|\frac{\delta p}{p}\right|^\up\sim
\left|\frac{\delta \rho}{\rho}\right|^\up \sim 
\left|\frac{\delta v}{v}\right|^\up.
%\sim \left|\frac{\delta Y_e}{Y_e}\right|^\up.
\label{eq:defx}
\end{equation}
Then  Eq.~(\ref{eq.p}) can be rewritten as
\begin{equation}
|x| \sim \frac{\displaystyle \left|\frac{\delta p}{p}\right|^\dw}
{ | \displaystyle f_\rho \mathrm{sgn}\left(\delta \rho^\up\right) -f_v\mathrm{sgn}\left(\delta v^\up \right) | },
\label{eq:finalx}
\end{equation}
where we ignore the terms of the upstream fluctuations of $p$ and $Y_e$ because of the relation, $|f_\rho|,\ |f_v| \gg |f_p|,\ |f_{Y_e}|$. Note that the signature in Eq.~(\ref{eq:finalx}) represents the {\it phase dependence} on perturbed flows. Since the flow fluctuates stochastically, we take average of the phase dependence in this study. As a result, we obtain the following relation for the amplitude of perturbations between pre- and post-shock wave; 
\begin{eqnarray}
%|x_{\rm ave}| &\equiv& \left< \sum_{\sigma_1, \sigma_2}  \frac{ \frac{\delta p}{p}\right|^\dw }{f_\rho \sigma_1 -f_v\sigma_2}  \right> \nonumber \\
|x_{\rm ave}| &\equiv& \left< \sum_{\sigma_1, \sigma_2} \frac{ \left|\frac{\delta p}{p}\right|^{\dw}  }{ | f_\rho \sigma_1 -f_v\sigma_2 | }  \right> \nonumber \\
&=& \frac{1}{2} \left|\frac{\delta p}{p}\right|^\dw  \left(\frac{1}{|f_\rho -f_v|}+ \frac{1}{|f_\rho +f_v|}\right),
\label{eq:xave}
\end{eqnarray}
where $\sigma_i\ (i=1,2)$ is either of $1$ or $-1$ and the brackets $<\cdots>$ mean the arithmetic mean. Finally, we insert $f_\mathrm{crit}^\dw$ into $\left|\delta p/p\right|^\dw$, then we get a simple relation for the critical amplitude of fluctuation at pre-shock ($f_\mathrm{crit}^\up$) and $f_\mathrm{crit}^{\dw}$,
\begin{equation}
f_\mathrm{crit}^\up = \frac{1}{2} \left(\frac{1}{|f_\rho -f_v|}+ \frac{1}{|f_\rho +f_v|}\right) f_\mathrm{crit}^{\dw}.
\label{eq:fcrit_pre_post}
\end{equation}
Note that we find that the dimensionless pre-factor in the right hand side of Eq.~(\ref{eq:fcrit_pre_post}) is less than unity, which means that $f_\mathrm{crit}^\up$ is smaller than $f_\mathrm{crit}^{\dw}$.

Here is a summary of Step 2. We use $r_\mathrm{sh}(M_r)$ and $M_\mathrm{PNS}(M_r)$ to evaluate the critical fluctuations at post-shock region ($f_\mathrm{crit}^{\dw}$) by Eq.~(\ref{fcrit_in_body}). By using the momentum flux balance between pre- and post- shock wave in Rankine-Hugoniot relation (Eq.~(\ref{eq:fcrit_pre_post})), we obtain the critical fluctuations at the pre-shock wave ($f_\mathrm{crit}^\up$). In the next step, we estimate $f_\mathrm{crit}^\mathrm{Si/O}$ from $f_\mathrm{crit}^\up$ by taking into account the amplification of perturbations during infall.

\subsection{Step 3: Amplification in supersonic accretion flows}\label{subsec.fluctu}

\begin{figure*}
\begin{tabular}{cc}
\begin{minipage}{0.45\hsize}
%\begin{center}
%\includegraphics[bb = 0 0 461 346, width = \columnwidth]{Mr_rho.png}
%\end{center}
\includegraphics[width = \columnwidth]{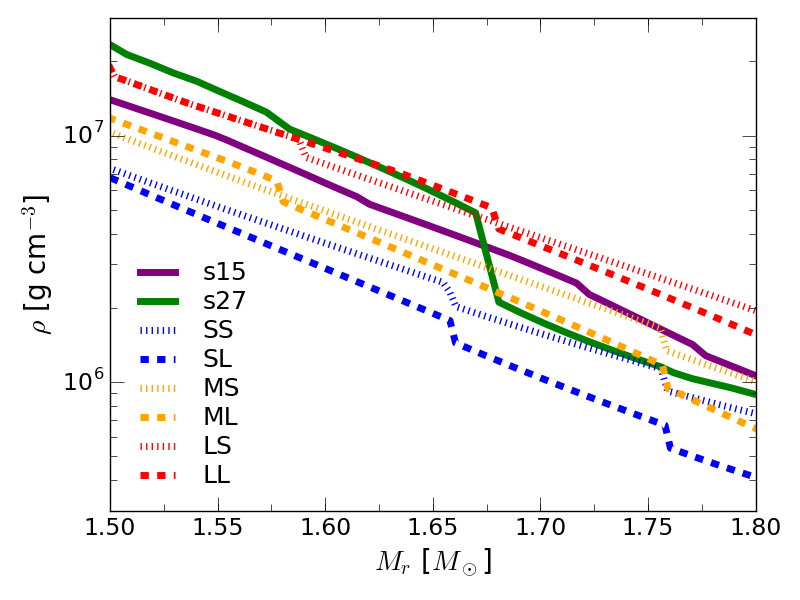}
\end{minipage} &
\begin{minipage}{0.45\hsize}
%\begin{center}
%\includegraphics[bb = 0 0 461 346, width = \columnwidth]{R_rho.png}
%\end{center}
\includegraphics[width = \columnwidth]{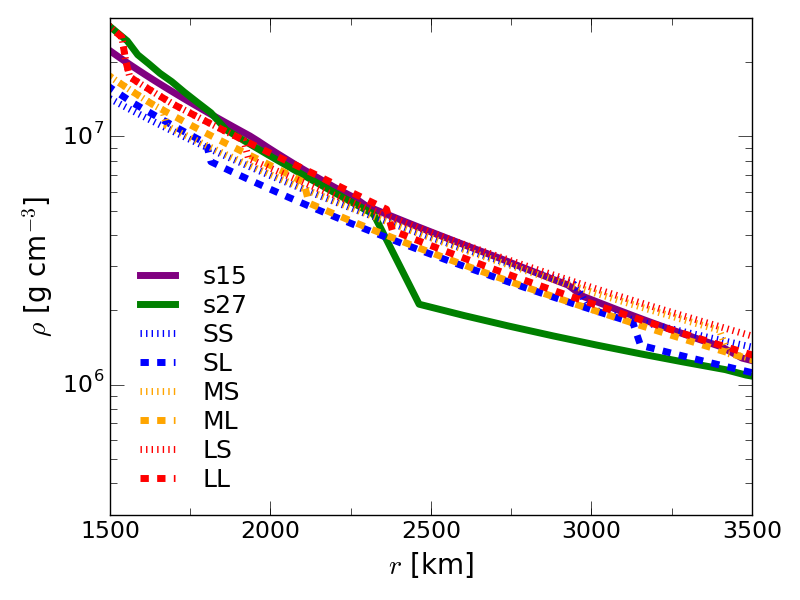}
\end{minipage}
\end{tabular}
\caption{The left and right panels show the density distribution before the onset of collapse with respect to the mass coordinate and radius for various progenitors, respectively. The thick and dashed lines are for the progenitors of \citetalias{WHW02} and \citetalias{YY}, respectively.}
\label{rhodist}
\end{figure*}

\begin{figure*}
\begin{tabular}{cc}
\begin{minipage}{0.42\hsize}
\includegraphics[width = \columnwidth]{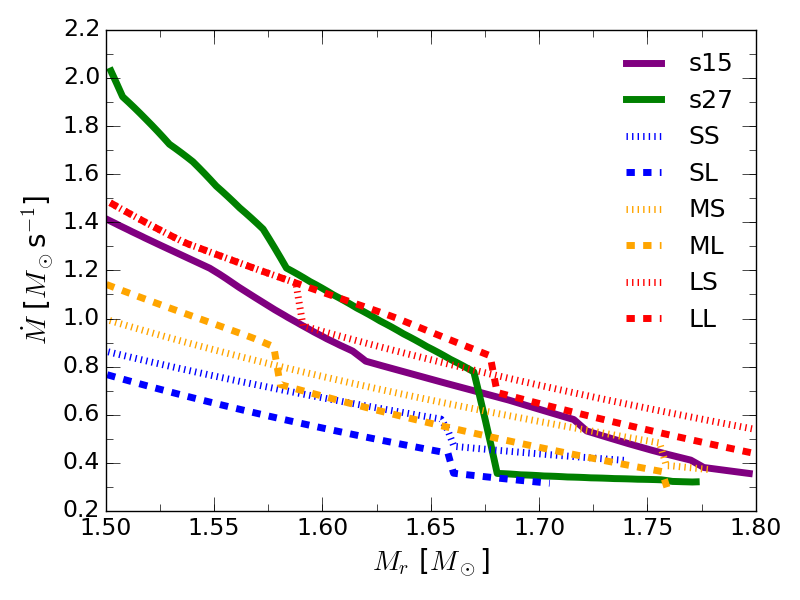}
\end{minipage} &
\begin{minipage}{0.42\hsize}
\includegraphics[width = \columnwidth]{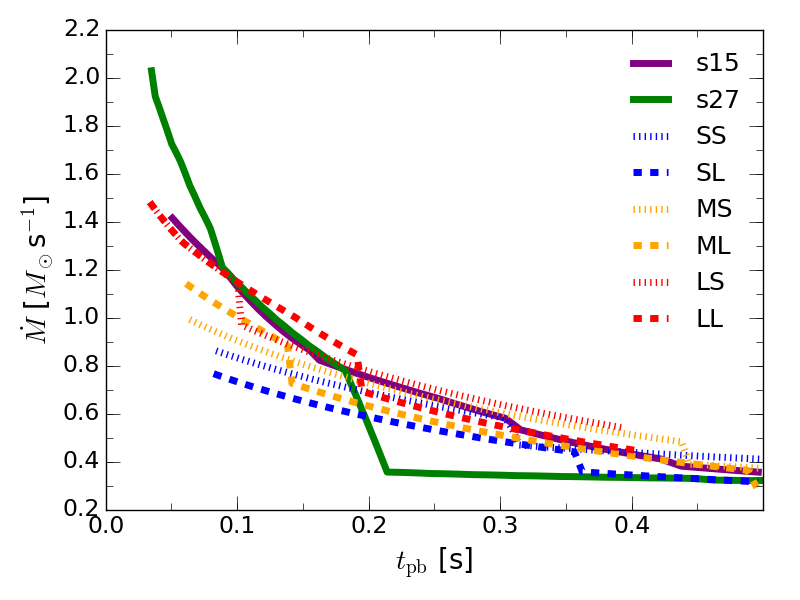}
\end{minipage} \\
%\begin{minipage}{0.42\hsize}
%\includegraphics[width = \columnwidth]{Mr-Lnu.png}
%\end{minipage} &
%\begin{minipage}{0.42\hsize}
%\includegraphics[width = \columnwidth]{tpb-Lnu.png}
%\end{minipage} \\
\begin{minipage}{0.42\hsize}
\includegraphics[width = \columnwidth]{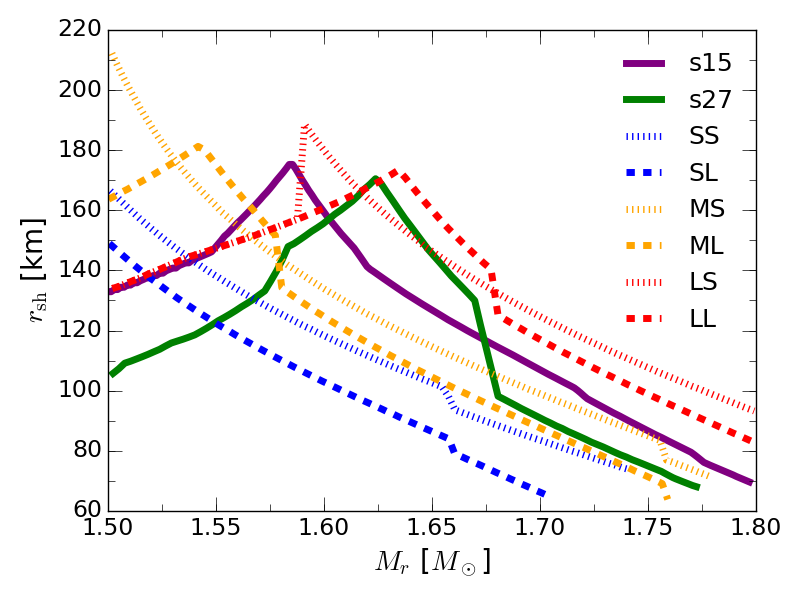}
\end{minipage} &
\begin{minipage}{0.42\hsize}
\includegraphics[width = \columnwidth]{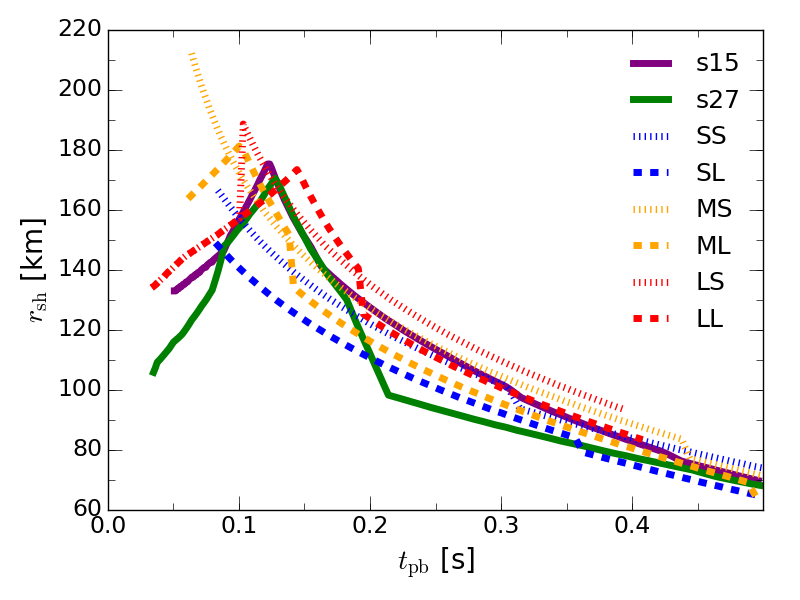}
\end{minipage}
\end{tabular}
\caption{The quasi-steady evolutions of $\dot{M}$ and $r_\mathrm{sh}$. The left panels show them as a function of $M_r$ while the right panels display the same quantities but as functions of $t_{\rm pb}$. The thick lines present the results for the s15 and s27 progenitors of \citetalias{WHW02} while the dashed ones are for the progenitors of \citetalias{YY}. The line types are the same as in Fig.~\ref{rhodist}.}
	\label{MLRsh}
\end{figure*}

\begin{table}
\centering
\caption{Core mass of progenitors. From left to right, the name of the model, the mass of the iron core, $M_\mathrm{Fe}$, the mass of the Ni layer and Si/S shells, $M_\mathrm{Ni+Si+S}$, and the reference are listed, respectively.}
\label{progenitors}
\begin{tabular}{ccccc}
\hline
Name& $M_\mathrm{Fe}$ & $M_\mathrm{Ni+Si+S}$  & Ref. \\
 & $[\mathrm{M_\odot}]$ & $[\mathrm{M_\odot}]$ & \\
\hline
s15 & 1.3 & 0.41 & WHW02 \\
s27 & 1.5 & 0.17 & WHW02\\
SS & 1.3 & 0.09 & YY16\\
SL & 1.3 & 0.18 & YY16\\
MS & 1.4 & 0.09 & YY16 \\
ML& 1.4 & 0.18 & YY16\\
LS & 1.5 & 0.09 & YY16\\
LL & 1.5 & 0.18 & YY16\\
\hline
\end{tabular}
\end{table}

Asymmetric fluctuations in accretion flows are, in general, growing in the supersonic regime. \citetalias{TY} investigated the growth of non-spherical perturbations during the infall onto the shock wave. They linearize the time-dependent hydrodynamic equations and then solve them by using a Laplace transform. They also analytically derived a scaling law of growth rate of perturbations as a function of radius, which is used to estimate $f_\mathrm{crit}^\mathrm{Si/O}$ from $f_\mathrm{crit}^{\up}$ in this paper. Below, we briefly review the derivation of the scaling law for the growth of fluctuations.

%  with generating waves of other modes even in linear level. According to \citetalias{TY}, density perturbations increase the amplitude in supersonic accretion flows with decreasing radius. We below review the derivation of the perturbation growth. 

%($\mathcal{M} \sim \mathcal{M}_0=\mathrm{const.}$)

\citetalias{TY} studied the time and spacial evolution of linear perturbations under steady spherical supersonic accretion flows\footnote{Note that \citetalias{TY} took into account the mixing modes between vorticity and pressure perturbations appropriately. In \citet{LG}, however, they imposed the irrotational condition in their analysis, which artificialy suppress a part of the mode coupling. This is one of the main reasons for the difference between two results.}. By using the fact that Mach number of a unperturbed radial flow ($\mathcal{M}$) changes slowly, they derived the following relation for the growth of non-radial ($\ell \ge 1$) density perturbation,
\begin{eqnarray}
\frac{\delta \rho}{\rho} (r) &=& \left. \frac{\delta \rho}{\rho} \right|_{r=r_*} \cos\left[\frac{\ell \ln(r/r_*)}{\sqrt{\mathcal{M}^2-1}}\right] \nonumber \\
&& + \frac{\ell \mathcal{M}^2}{\sqrt{\mathcal{M}^2-1}} \left. \frac{\delta v_\perp}{v_r} \right|_{r=r_*} \sin\left[\frac{\ell \ln(r/r_*)}{\sqrt{\mathcal{M}^2-1}}\right],
\end{eqnarray}
where $r_*$ denotes an arbital radius.
%which follows from the linearized conservation laws of mass and momentum.
 The sinusoidal functions give the oscillating pattern in space. Since $\mathcal{M}>1$ in supersonic flows and $\ell \ge 1$, the second term dominates the first one for $\delta \rho/\rho$ and $\delta v_\perp/v_r$ being the same order at $r=r_*$.

% $\delta \rho/\rho|_{r=r_*} \sim \delta v_\perp/v_r|_{r=r_*}$
% except where the value in the brackets of $\sin$ becomes a multiple of $\pi$, and the maximal magnitude is given by $ \ell \mathcal{M}_0^2/\sqrt{\mathcal{M}_0^2-1} \times (\delta v_\perp/v_{r0})|_{r_0}$.
We note that the Mach number gradually increases in accretion flows: e.g., $\mathcal{M} \propto r^{-(5-3\gamma)/4}$, where $\gamma$ is the ratio of specific heats, while $\delta v_\perp/v_{r}$ does not grow with radius \citepalias{TY}. In this case, the growth of the maximal magnitude, or in other words, the growth of the envelope of the sinusoidal oscillation in space is roughly given by
\begin{equation}
\label{envelope_rho}
\frac{\delta \rho}{\rho}(r) \sim \ell \left(\frac{r}{r_c}\right)^{-(5-3\gamma)/4} \left. \frac{\delta v_\perp}{v_{r}}\right|_{r=r_c} ,
\end{equation}
where we choose $r_* =r_c$ (the radius of the trans-sonic point) in this expression.
% The left hand side of Eq.~(\ref{envelope_rho}) is interpreted as the envelope size of oscillating perturbation pattern.
 Similarly, the growth of pressure perturbation can be also written as
\begin{equation}
\label{envelope}
\frac{\delta p}{p} (r) \sim \gamma \ell \left(\frac{r}{r_c}\right)^{-(5-3\gamma)/4} \left. \frac{\delta v_\perp}{v_{r}}\right|_{r=r_c},
\end{equation}
which holds well for $\ell \gtrsim 4$ in a Bondi accretion flow \citepalias{TY}. By using Eq.~(\ref{envelope}), we can obtain the following relation between $f_\mathrm{crit}^\mathrm{Si/O}$ and $f_\mathrm{crit}^\up$,
%%  the envelope size of pressure perturbation reaches $f_\mathrm{crit}^\up$ at $r=r_\mathrm{sh}$ when the following equation is satisfied 
%% \begin{equation}
%% f_\mathrm{crit}^\up \sim \gamma \ell \left(\frac{r_\mathrm{sh}}{r_c}\right)^{-(5-3\gamma)/4} \left. \frac{\delta v_\perp}{v_{r0}}\right|_{r_0},
%% \end{equation}
%% from which we obtain the pressure perturbation given at $r=r_c$ that grows to $f_\mathrm{crit}^\up$ at $r=r_\mathrm{sh}$:
\begin{equation}
\label{fcritini_in_body}
f_\mathrm{crit}^\mathrm{Si/O} \equiv \left. \frac{\delta p}{p}\right|_{r=r_c} \sim f_\mathrm{crit}^\up\left(\frac{r_\mathrm{sh}}{r_c}\right)^{(5-3\gamma)/4}.
\end{equation} 
Hereafter we set $\gamma=4/3$ for all the shells, although $\gamma$ could be varied among shells and even changed during infall by the deleptonization and nuclear burning (see also Sec.~\ref{nonlineareffects}). Since the radius of the trans-sonic point $r_c$ is not varied among progenitors, we approximately set $r_c = 10^8$~cm in this study. We interpret the pressure perturbation at $r=r_c$ as the initial fluctuation in shells before the collapse $f_\mathrm{crit}^\mathrm{Si/O}$, assuming that the fluctuations do not grow nor decrease in the subsonic regions.

Equation~(\ref{fcritini_in_body}) shows an important fact that $f_\mathrm{crit}^\mathrm{Si/O}$ becomes smaller with decreasing $r_\mathrm{sh}$ for a fixed $f_\mathrm{crit}^\up$. This is a consequence of the amplification of progenitor asymmetry during infall. On the contrary, smaller $r_\mathrm{sh}$ leads to larger $f_\mathrm{crit}^\up$ (see Eq. (14)), which means that the two effects compete against each other with respect to the change of $r_\mathrm{sh}$. We find that $100 \lesssim r_\mathrm{sh} \lesssim 200$km is the threshold region. In the region $r_\mathrm{sh} \gtrsim 200$km, the latter effect becomes dominant to determine $f_\mathrm{crit}^\mathrm{Si/O}$, while the former becomes dominant in $r_\mathrm{sh} \lesssim 100$km (see Sec.~\ref{effrsh} in more detail).
% As we shall show below, the shock waves evolve $r_\mathrm{sh}\lesssim 200$~km in our quasi-steady model. In this region, the progenitor amplification dominates the radial dependence of $f_\mathrm{crit}^\up$.

% i.e., reduces $f_\mathrm{crit}^\mathrm{Si/O}$ with decreasing $r_\mathrm{sh}$.
% {\bf Although this is one of important results in this paper, we give a caution before proceeding to analyze our results. Since the non-linear mode coupling may suppress the growth of perturbation, we would overestimate the effect of the amplification of fluctuations during infall. This is the biggest weakness of our method. We will modify our models by comparing more sophisticated models or numerical simulations in the forthcoming paper.}

%%%%%%

\section{Progenitors}\label{sec.prog}

In this paper we employ $15$ and $27 M_\odot$ progenitor models from realistic stellar evolution in \citet{WHW02} (s15 and s27), and six parametrically generalized progenitors developed by \citetalias{YY} (SS, SL, MS, ML, LS and LL). We first apply our model to s15 and s27 in order to calibrate the free parameters of our model (see Sec.~\ref{subsec.quasisteady} for more details). Those calibrated parameters are employed for the rest of other progenitors. For the parameterized progenitors, \citetalias{YY} categorized various progenitors in the literature into some groups, which is useful to cover many types of progenitors than employing models from stellar evolution calculations. We show the mass of the iron core, $M_\mathrm{Fe}$, and the mass of the Ni layer and Si/S shells, $M_\mathrm{Ni+Si+S}$, in Table~\ref{progenitors}\footnote{In the table, $M_\mathrm{Fe}$ and $M_\mathrm{Ni+Si+S}$ are defined as follows. For the s15 and s27 progenitors, they are the masses of the regions dominated by irons and by nickel, silicon and sulfur, respectively. On the other hand, for \citetalias{YY} progenitors, they correspond to the masses of the nuclear statistical equilibrium and quasi-statistical equilibrium regions, respectively.}. The radial profile of density distribution at the onset of collapse is displayed in Fig.~\ref{rhodist}.

Note that we do not consider electron-capture or low-mass-iron CCSN progenitors in this study. In principle, our model can be applied to them. However they would achieve the shock revival without any aid from progenitor asymmetries (\citet{2006A&A...450..345K,2017ApJ...850...43R}, and see also a recent review in \citet{2016PASA...33...48M}).

\section{Results}\label{sec.results}

\begin{figure}
%	\begin{center}
%		\includegraphics[bb = 0 0 461 346, width=\columnwidth]{tpb-Mpns.png}
%	\end{center}
\includegraphics[width=\columnwidth]{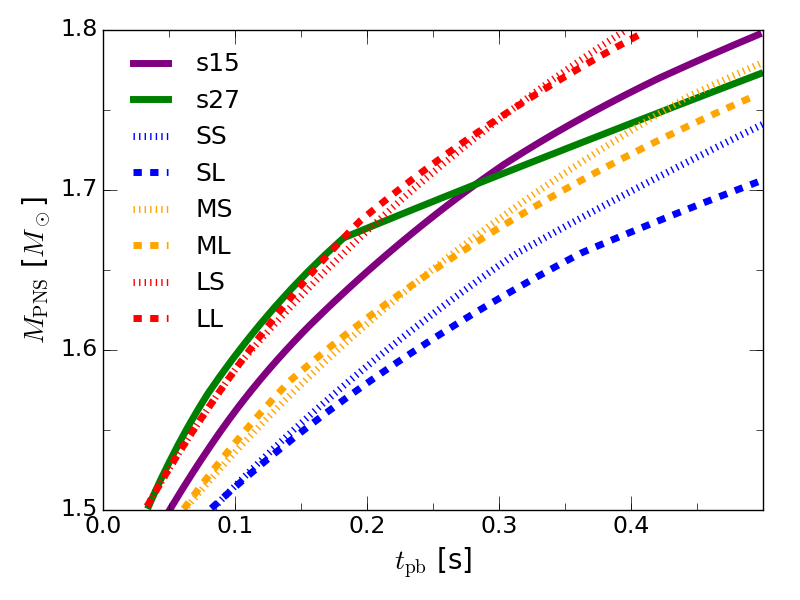}
\caption{The time evolution of $M_\mathrm{PNS}$ as a function of $t_{\rm pb}$ in our fiducial model. The line types are the same as in Fig.~\ref{rhodist}.}
\label{evo.Mpns}
\end{figure}

\subsection{Quasi-steady models}\label{subsec:result_quasisteady}
Figure~\ref{MLRsh} shows the result of $\dot{M}$ and $r_\mathrm{sh}$ obtained by our quasi-steady model. We display them as a function of $M_r$ ($t_{\rm pb}$) on the left (right) panels. For the mass shell of $1.5\lesssim M_r/M_\odot \lesssim 1.8$, we find steady solutions, which indicates that the neutrino luminosity in our model are smaller than the critical neutrino luminosity. This is qualitatively consistent with the fact that stagnant shock waves do not revive only with the neutrino-heating process in 1D \citep{RJ00,Mezzacappa01,Liebendorfer05,Sumiyoshi,Suwa}. 
%More importantly, the shock trajectories in our model are quantitatively consistent with the result of numerical simulations. These are owing to two additional prescriptions in our model; establishing a model to define $\dot{M}$, $M_\mathrm{PNS}$ and $L_\nu$ as a function of $M_r$ under the given the progenitor structure; calibrating other free parameters to reproduce the simulation results. In fact, the clasical quasi-steady models with light-bulb approximation are not capable of reproducing simulation results quantitatively (see e.g., \citet{Yamasaki06}).

% Our two efforts, 

% Figure~\ref{MLRsh} 
%improve the semi-analytica model
%owing to 
% by 

%Owing to the detailed model for the connection 
% Owing to the calibration of our free parameters, our model 
%These are owing to the calibration of free parameters.
% that our quasi-steady model is well calibrated by 
% which is consistent with the results of detailed spherically symmetric simulations (see also Appendix~\ref{sec.calb}). 

Progenitors with a larger iron core tend to have larger $\dot{M}$, which is caused by the compact envelope shown in the left panel in Fig.~\ref{rhodist}. On the other hand, the time evolution of $r_\mathrm{sh}$ is not a monotonic function of the zero-age main sequence (ZAMS) mass of progenitor. The increase of $\dot{M}$ tends to kinematically push the shock wave back. On the other hand, larger $\dot{M}$ produces higher accretion components of neutrino luminosity, which, in contrary, works to push the shock wave outward as a result of increase of neutrino heatings. Two effects compete against each other and then smear out the monotonic correlation between $r_\mathrm{sh}$ and the ZAMS mass of progenitor.

%, the time evolution of $r_\mathrm{sh}$

% which are the mass accretion rate ($\dot{M}$), total neutrino luminosity ($L_\nu$) and shock radius ($r_\mathrm{sh}$) for the progenitors listed in Table~\ref{progenitors}. The left panels present these quantities as functions of $M_r$ with $1.5$~$\mathrm{M_\odot} \le M_r \le 1.8$~$\mathrm{M_\odot}$ while the right ones give them as functions of the post-bounce time ($t_{\rm pb}$) defined by $t_{\rm pb} = t_\mathrm{infall}-t_\mathrm{infall}(M_r=1.42\ \mathrm{M_\odot})$, where we calibrated as the bounce time coincides with when when the mass shell of $M_r=1.42M_\odot$ hits the shock wave. We also plot the evolution of $M_\mathrm{PNS}$ as a function of $t_{\rm pb}$ in Fig.~\ref{evo.Mpns}.
%\footnote{Note that the choice of $t_\mathrm{lag}$ does never essentially affect our results nor discussion.} 
%The top panels in Fig.~\ref{MLRsh} shows the result of the mass accretion rate $\dot{M}$. The initial mass accretion rate appears to be larger for more dense progenitors with larger $\rho$. Progenitors with a larger iron core produce larger $\dot{M}$, which is caused by the compact envelope shown in the left panel in Fig.~\ref{rhodist}. It decreases with time and suddenly drops at shell interfaces.

The progenitor dependence of $\dot{M}$ results in different time evolutions of the growth of $M_\mathrm{PNS}$ as shown in Fig.~\ref{evo.Mpns}. The PNS mass increases faster for the progenitors with higher mass accretion rates. Indeed, the mass shell of $M_r=1.8M_\odot$ can fall onto the shock wave by $t_{\rm pb}=0.5$s for compact progenitors (s15, LS and LL) but not for other light progenitors. $M_\mathrm{PNS}$ is the primary term to determine the gravitational binding energy on CCSNe core, which affects both a solution of quasi-steady model and the critical fluctuation of shock revival. Note that, since $\dot{M}$ is well calibrated in our model, the time evolution of $M_\mathrm{PNS}$ is also quantitatively consistent with numerical simulations.

\subsection{Required progenitor asymmetries for the shock revival}\label{subsec:reqasym}

\begin{figure*}
\begin{tabular}{cc}
\begin{minipage}{0.48\hsize}
\includegraphics[width = \columnwidth]{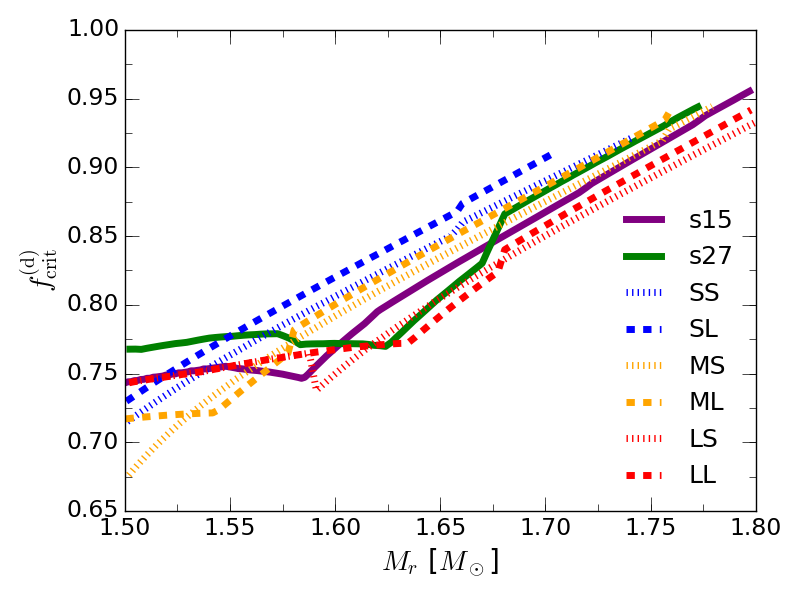}
\end{minipage} &
\begin{minipage}{0.48\hsize}
\includegraphics[width = \columnwidth]{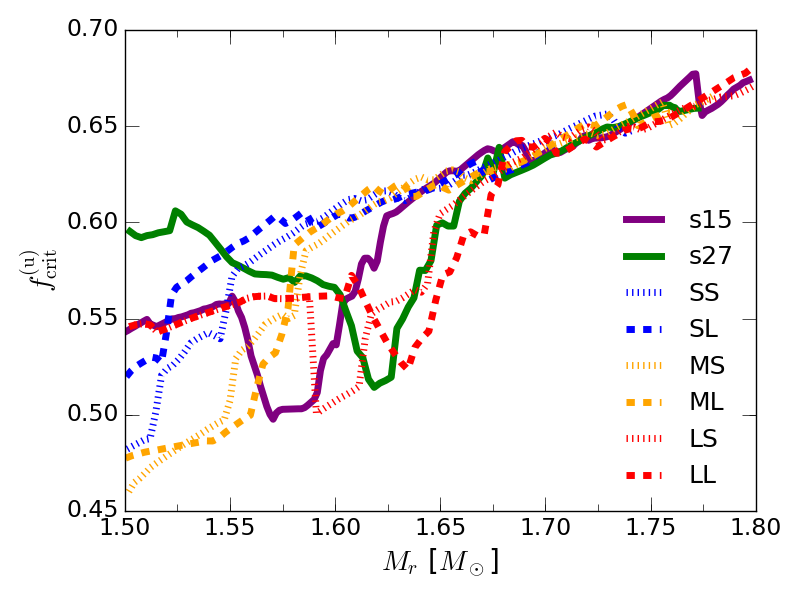}
\end{minipage}
\end{tabular}<
\caption{The results of $f_\mathrm{crit}^{(d)}$ and $f_\mathrm{crit}^{(u)}$ for our fiducial model. The line types are the same as in Fig.~\ref{rhodist}.}
\label{fig.fcrit}
\end{figure*}

We now turn our attention to progenitor asymmetries.
% to give an impact for the shock revival.
%\subsubsection{Radial profiles of $f_\mathrm{crit}^{\dw}$, $f_\mathrm{crit}^{\up}$} and $f_\mathrm{crit}^\mathrm{Si/O}$\label{sec.fcrit_body}
The left panel in Fig.~\ref{fig.fcrit} shows $f_\mathrm{crit}^{\dw}$ as a function of $M_r$. One of the remarkable features of $f_\mathrm{crit}^{\dw}$ is that it almost monotonically increases with $M_r$. This is caused by the decrease of the shock radius and the increase of the PNS mass with $M_r$, both of which make the shock wave sink into deeper gravitational potential.
% Larger fluctuations are required for the shock wave to escape from the core.
% Note that $f_\mathrm{crit}^\dw$ can decrease in $M_r \lesssim 1.6M_\odot$ for s15, s27, LL and LS progenitors, for which the shock initially increases to the peak in the early phase.

Another notable feature is that the profiles of $f_\mathrm{crit}^{\dw}$ are quite similar for the employed progenitors. This is simply because the difference of the shock radii ($\sim10^{6\mathrm{-}7}$~cm) is so small to be reflected to $f_\mathrm{crit}^{\dw}$. In fact, the susceptibility of $f_\mathrm{crit}^{\dw}$ to a shift of the shock radius, $\delta r_\mathrm{sh}$, is estimated by using Eq.~(\ref{fcrit_in_body}) as:
\begin{equation}
\label{sensibility}
\frac{\delta f_\mathrm{crit}^{\dw}}{f_{\mathrm{crit},0}^{\dw}} = \frac{\delta r_\mathrm{sh}}{10^{8}\ \mathrm{cm} -r_{\mathrm{sh},0}}\sim \frac{10^{6\mathrm{-}7}\ \mathrm{cm}}{10^8 \ \mathrm{cm}-10^7\ \mathrm{cm}}\sim 10^{-2}\mathrm{-}10^{-1},
\end{equation}
where the letters with a subscript $0$ denote some baseline values. Thus, $f_\mathrm{crit}^{\dw}$ changes by only $1\%$ when the shock radius changes by $\sim 10\%$.
% Therefore, $f_\mathrm{crit}^{\dw}$ is not sensitive to the difference of $r_\mathrm{sh}$ as also shown in Appendix~\ref{appModel}.

%The corresponding pressure fluctuation for the upstream side, $f_\mathrm{crit}^\up$, is shown in the right panel in Fig.~\ref{fig.fcrit}.

On the other hand, we find that $f_\mathrm{crit}^\up$ is roughly $\sim 30\%$ smaller than $f_\mathrm{crit}^\dw$, which is shown in the right panel in Fig.~\ref{fig.fcrit}. We also find that $f_\mathrm{crit}^\up$ varies with $M_r$ more than $f_\mathrm{crit}^{\dw}$ does. This is mainly attributed to the fact that the connection of perturbed quantities between pre- and post-shock wave depends sensitively on jump condition at the shock surface.
% The change of $r_\mathrm{sh}$ is a primary factor for the variation since $r_\mathrm{sh}$ dictates the strength of shock wave in CCSNe core at least for quasi-steady regime (stronger shock wave for smaller $r_\mathrm{sh}$). Indeed, some dips of $f_\mathrm{crit}^up$ seen in the right panel in Fig.~\ref{fig.fcrit} for s15, s27 and LS correlate with the sudden change of $r_\mathrm{sh}$ (see the left and bottom panel in Fig.~\ref{MLRsh}).
 Roughly speaking, however, that $f_\mathrm{crit}^\up$ grows with $M_r$ by dragging the property of $f_\mathrm{crit}^\dw$ in particular for the outer mass shell ($M_r \gtrsim 1.65 M_\odot$).

\begin{figure}
\begin{tabular}{c}
\begin{minipage}{1\hsize}
%\begin{center}
%\includegraphics[bb = 0 0 461 346, width = \columnwidth]{Mr-fcritSiO.png}
%\end{center}
\includegraphics[width = \columnwidth]{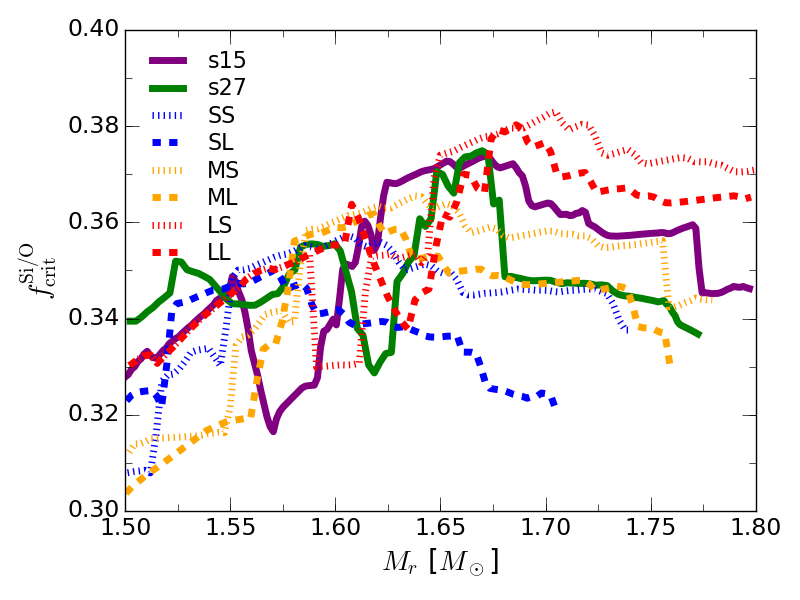}
\end{minipage} 
\end{tabular}
\caption{The results of $f_\mathrm{crit}^\mathrm{Si/O}$ for our fiducial model. The line types are the same as in Fig.~\ref{rhodist}.}
\label{fig.fcritSiO}
\end{figure}

The result of $f_\mathrm{crit}^\mathrm{Si/O}$ is shown in Fig.~\ref{fig.fcritSiO}, which is the most important outcome in this paper. The amplification of progenitor asymmetries during infall reduces a factor $\sim 2$ of the critical amplitude from $f_\mathrm{crit}^{\up}$ and results in $0.3~\lesssim~f_\mathrm{crit}^\mathrm{Si/O}~\lesssim~0.4$ for the displayed mass range. We also find that $f_\mathrm{crit}^\mathrm{Si/O}$ roughly increase with $M_r$ for $M_r< 1.7 M_\odot$ and then being flat for the larger mass shell. The trend is along roughly with $f_\mathrm{crit}$, but the flat profile for the larger mass shell is attributed by the decrease of shock radius, which results in prolonging the supersonic accretion regime and promoting the growth of fluctuations during infall.

\subsection{Importance of progenitor asymmetry to shock revival}\label{sec.cruciality}
As shown above, we obtain the progenitor asymmetry required at the pre-supernova stage for shock revival as $0.3~\lesssim~f_\mathrm{crit}^\mathrm{Si/O}~\lesssim~0.4$ for the employed progenitors. Below, we apply our results to diagnose the importance of progenitor asymmetry to shock revival.

%Our result would be robust within the uncertainties of free parameters (see in Appendix~\ref{appModel})
%although there remains some limitations for the applicability of our model, which are discussed in Sec.~\ref{sec.limitation}.

% Note that this range is also valid even for other model parameter sets listed in Table \ref{models} (models B-I) (See Appendix~\ref{appModel}).

% The lower limit can be $\sim???$ if we take into account uncertainties of our models as discussed in the next section. 
The current stellar evolution models predict that the fluctuations in the Si/O shells at the pre-supernova stage would be roughly less than $10 \%$, which means that $f_\mathrm{crit}^\mathrm{Si/O}$ is roughly three times larger than our prediction. Thus, we reach a conclusion that the progenitor asymmetry in realistic stellar models does not have enough power to launch a shock revival.

Another important finding in this paper is that $f_\mathrm{crit}^\mathrm{Si/O}$ is smaller in the inner mass shells, since $f_\mathrm{crit}^\mathrm{Si/O}$ increases with $M_r$ albeit the minor dependence. This means that progenitor asymmetries in the vicinity of an iron core would give an impact of shock dynamics even if they are not enough to drive shock revival at this time. They would play a supplementary but important role for shock revival through enhancing the turbulence and convection in the shocked region, which is another possible way to give an impact to the shock dynamics, though. Thus, we may need to care not only the amplitude of progenitor asymmetries but also their {\it locations} for the comprehensive understanding of the role of progenitor asymmetry to shock revival.

% Note that our results can be applied, in principle, for electron-capture or low-mass-iron CCSNe progenitors. It should be noted, however that they would achieve the shock-revival without any aid from progenitor asymmetries \citet{2015MNRAS.453..287M,2017MmSAI..88..288M,2017ApJ...850...43R}.

% The fluctuations at smaller radii could be more important than ones at larger radii.

\section{Limitations of the current study}\label{sec.limitation}
%In this paper we developed an elaborate semi-analytica model to study the importance of progenitor asymmetry to shock revival. However, since there still remains some uncertainties in our model, below we discuss how they affect our conclusions. 
In this section, we give several important limitations in our model. We also discuss how they affect our conclusion.
%give a few cautions for the results presented in this paper.
% it still includes some uncertainties 
% there remains uncertainties 
%Although it is a robust conclusion that progenitor fluctuations cannot play a primary role to revive a stalled shock wave in the neutrino heating mechanism, there are some important cautions for the results presented in this paper. We describe them below and discuss how our results can be changed.

\subsection{Uncertainties of $f_\mathrm{crit}^\mathrm{Si/O}$ as pre-collapse perturbations}\label{precolapertu}
We use $f_\mathrm{crit}^\mathrm{Si/O}$ to diagnose the importance of progenitor asymmetry to shock revival by comparing it with a canonical amplitude of progenitor asymmetry in stellar evolution. Strictly speaking, however, the two asymmetric quantities are not identical since the perturbation may grow or decay during the subsonic infall phase. At the moment, unfortunatelly, there are no analytic or semi-analytic methods to quantify the change of perturbation in the subsonic phase. The solutions would be given by the systematic numerical simulations for the collapsing phase of CCSNe including progenitor asymmetries (see e.g., 2D simulations in \citet{MJ14}). The systematic study would reveal some statistical properties of the growth/decay of perturbations, which would be useful for our semi-analytical model. We will address the issue in the forthcoming paper.

\subsection{Uncertainties of shock radius}\label{effrsh}
Our quasi-steady model is well calibrated by the result of numerical simulations. However, any computational method even for detailed numerical simulations is an approximation of reality, which indicates that our obtained shock evolution is different from the reality as well. Hence, it is worth estimating how the uncertainty of the shock radius affects our results.

The expansion of a shock wave can bring two opposite effects to $f_\mathrm{crit}^\mathrm{Si/O}$. As can be seen in Eq.~(\ref{fcritini_in_body}), $f_\mathrm{crit}$ is reduced with increasing $r_\mathrm{sh}$ due to being less bound by gravity, while the larger $r_\mathrm{sh}$ shortened supersonic accretion and then suppresses the amplification of fluctuations during infall. The change rate of $f_\mathrm{crit}^\mathrm{Si/O}$ for a small displacement of the shock, $\delta r_\mathrm{sh}$, is given by Eqs.~(\ref{fcrit_in_body}) and (\ref{fcritini_in_body}) as follows:
\begin{equation}\label{deltaf}
\frac{\delta f_\mathrm{crit}^\mathrm{Si/O}}{f_{\mathrm{crit},0}^\mathrm{Si/O}} \sim \left( - \frac{r_{\mathrm{sh},0}}{10^8\ \mathrm{cm}-r_{\mathrm{sh},0}} + \frac{5-3\gamma}{4} \right)  \frac{\delta r_\mathrm{sh}}{r_{\mathrm{sh},0}} \equiv \zeta \frac{\delta r_\mathrm{sh}}{r_{\mathrm{sh},0}},
\end{equation}
where a subscript $0$ denotes the quantities without shock displacement. Note that we ignore the conversion between $f_\mathrm{crit}^{\up}$ and $f_\mathrm{crit}^{\dw}$ in the above estimation since it can not be described explicitly as a function of $\delta r_\mathrm{sh}$ (see Eq.~(\ref{eq:fcrit_pre_post})). Although the conversion factor quantitatively affects the following discussion, Eq.~(\ref{deltaf}) is enough to catch the essence of the argument. In Eq.~(\ref{deltaf}), two opposite effects can be seen as two competing terms in the parentheses: $f_\mathrm{crit}^\mathrm{Si/O}$ is increased ($\zeta > 0$) or decreased ($\zeta <0$) for a given shock expansion $\delta r_\mathrm{sh}>0$. Below, we discuss the property of $\zeta$ in detail.

\begin{figure}
\includegraphics[width=\columnwidth]{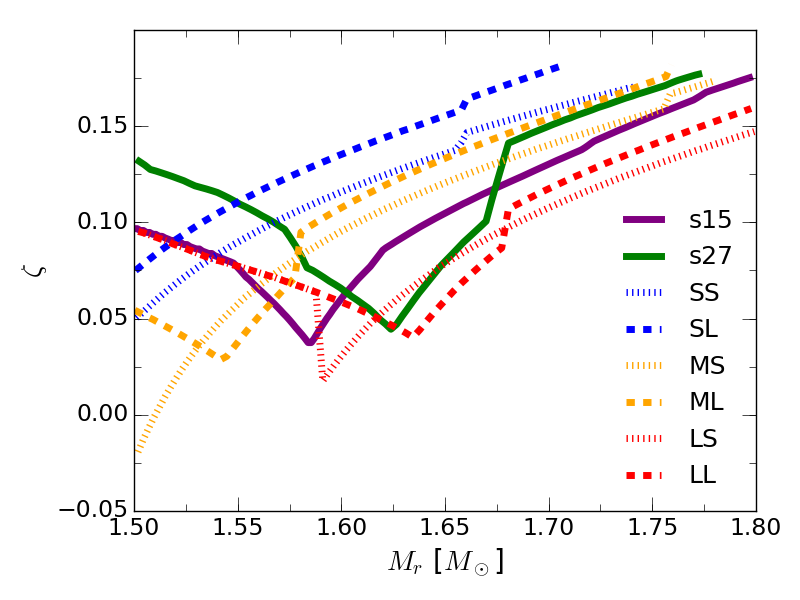}
\caption{The profile of $\zeta$ as a function of $M_r$. The line types are the same as in Fig.~\ref{rhodist}.}
\label{fig.zeta}
\end{figure}

\begin{figure}
\begin{tabular}{c}
\begin{minipage}{1\hsize}
%\begin{center}
%\includegraphics[bb = 0 0 461 346, width = \columnwidth]{fix_Mr-fcritSiO.png}
%\end{center}
\includegraphics[width = \columnwidth]{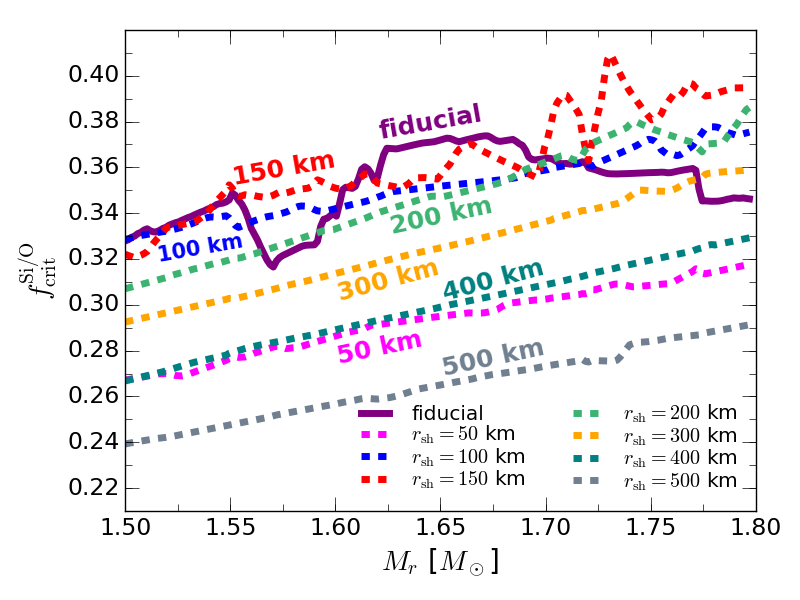}
\end{minipage} 
\end{tabular}
\caption{The profile of $f_\mathrm{crit}^\mathrm{Si/O}$ of the s15 progenitor but the shock radius is artificially fixed to some radius throughout the mass range. For comparison, the result for our fiducial model is also plotted by the thick purple line.}
\label{fig.100-500}
\end{figure}

Figure~\ref{fig.zeta} shows $\zeta$ as a function of $M_r$ for the employed progenitors. As seen in the figure, $\zeta$ is between 0.05 and 0.15 except for the very early phase of MS model. Hence, for most of the regime, $f_\mathrm{crit}^\mathrm{Si/O}$ increases by $\sim1 \%$ when the shock radius becomes larger by $\sim 10 \%$ than in our fiducial model. This is due to the fact that, for small shock radii, the suppression of amplification by shortened supersonic accretion dominates the rate of change, i.e., the second term in the middle equation in Eq.~(\ref{deltaf}) overwhelms the first one. We note, however, that the first term becomes dominant, i.e., $\zeta$ transits to be negative, as $r_\mathrm{sh}$ approaches $10^8$~cm. The turning point can be estimated by solving $\zeta = 0$ with $\gamma=4/3$ and we obtain $r_\mathrm{sh} = 200$~km. This non-monotonic behavior of $f_\mathrm{crit}^\mathrm{Si/O}$ is illustrated in Fig.~\ref{fig.100-500}, where $f_\mathrm{crit}^\mathrm{Si/O}$ of the s15 progenitor is plotted for several constant shock radii ($50 < r_\mathrm{sh} < 500$~km). Note that the actual transition occurs between $100 \lesssim r_\mathrm{sh} \lesssim 200$~km. The quantitative deviation is due to the conversion factor from $f_\mathrm{crit}^{\dw}$ to $f_\mathrm{crit}^{\up}$.

The modification of $f_\mathrm{crit}^\mathrm{Si/O}$ due to the uncertainty of $r_\mathrm{sh}$ is at most $\sim20 \%$. Even for $r_\mathrm{sh} = 500$km, $f_\mathrm{crit}^\mathrm{Si/O}$ is larger than $\sim 0.24$, which means that we still need a larger progenitor asymmetry than results by stellar evolutions. Therefore, we can conclude that the uncertainties of shock wave do not affect our conclusions. 
%Note that $L_\nu$ approaches to the crtical luminosity for the solution with increasing $r_\mathrm{sh}$, and then the unperturbed state would become unstable to a radial overstabilization mode \citep{Fernandez12} even in spherically symmetric conditions. Therefore, the shock revival would be happening before $L_\nu$ reaches the crtical luminosity even without any aid by progenitor asymmetries.

% As shown in the figure, $f_\mathrm{crit}^\mathrm{Si/O}$ increases with $r_\mathrm{sh}$ until $r_\mathrm{sh}$ reaches $2\times10^7$~cm, whereas $f_\mathrm{crit}^\mathrm{Si/O}$ decreases with $r_\mathrm{sh}$ beyond the turning point. Within the range of $0.5\times10^7~\mathrm{cm}\le r_\mathrm{sh}\le 5\times10^7~\mathrm{cm}$, $f_\mathrm{crit}^\mathrm{Si/O}$ varies from $\sim0.25$ to $\sim0.6$. 

\subsection{Non-linear effects in the growth of asymmetry during infall}\label{nonlineareffects}
We use a scaling law (Eq.~(\ref{fcritini_in_body})) to estimate the growth of asymmetry during infall, which was derived based on the linear analysis of \citetalias{TY}. In reality, however, non-linear effects can not be neglected once the asymmetric fluctuation grows sufficiently and reaches the order of unperturbed flows. Indeed, we find that $f_\mathrm{crit}^{\up}$ is larger than $\sim 0.5$ and even reaches $\sim 0.7$ in particular for the outer envelope (see the right panel in Fig.~\ref{fig.fcrit}). The magnitude is enough large to require considerations for how non-linear effects affect our conclusion.

Although the non-linear effects of growing asymmetries during infall has not been well studied (but see \citet{Couch3,Muller17} for numerical simulations), one of the major effects would be the non-linear saturation due to the mode coupling. If the saturation really occurs, the growth of fluctuations would be suppressed. This means that $f_\mathrm{crit}^\mathrm{Si/O}$ in our study is underestimated, i.e., our result is conservative.
% and being not necessary to be revised.

Another deficit in our model is that we do not take into account the effect of nuclear burning which takes place, in general, during infall (see e.g., \citet{Yamamoto13}). $\gamma$ is also changed as a result of nuclear burning, which also cause of errors in our estimation. More importantly, the nuclear burning couples with matter fluctuations non-linearly and may cause the enhancement of the fluctuation. Although it may affect our conclusion (since $f_\mathrm{crit}^\mathrm{Si/O}$ presented in this paper would be overestimated by the ignorance of effect of nuclear burning), it is dificult to include the effect in our semi-analytical model since the detailed nuclear network calculations would be required. This is one of the major uncertainties in this study which should be addressed in the future work.
% To address this issue, we need to develop more elaborate semi-analytica approach or the hybrid method of semi-analytic method and numerical simulations, which is on our to do list of future investigations.

%quantitatively

% requirement since the amplification rate could be further enhanced by nuclear burning which results in the 

% related to the lack of input physics in our method, which is

%should be suppressed 

\subsection{Comparison with other criteria for shock revival}\label{efffcrit}
%\subsection{Uncertainties of the critical condition for the shock revival}\label{efffcrit}
We apply a criterion (Eq.~(\ref{fcrit_in_body})) to judge the condition of shock revival. The simple criterion was derived based on the result of the semi-dynamical approach developed by \citetalias{CritFluc}, and the applicability was confirmed by comparing to the result of an axisymmetric simulation with light-bulb neutrino transport. However, various approximations are implied even in the numerical simulation and the uncertainties may potentially change our conclusion as well. Thus, it is worth to see other criteria. The comparison would brush up our model although these improvements will be made in our future work.

%In addition to this, the critical value would depend on the geometry of fluctuations which was not taken into account in the creterion.

% discuss the applicability of them

% proposed by different groups to measure the importance of progenitor asymmetries for shock revival.

The critical Mach number $<\mathcal{M}_a^2>$ is one of the interesting criteria to measure the impact of progenitor asymmetry, which was originally proposed by \citet{MJ14} and has been used to analyze numerical simulations (see e.g., \citet{Summa}). The criterion was made based on the idea that the violent aspherical motions in the post-shock flows can reduce the critical luminosity. \citet{MJ14,Summa} found that $<\mathcal{M}_a^2>\sim 0.3$ is a threshold value for the runaway shock evolution, which results in reducing the critical luminosity $\sim 25 \%$ compared to spherically symmetric case.

If we adopt the critical Mach number to determine the condition for shock revival, the required pressure fluctuation in the post-shock flows can be roughly estimated by $\delta p /p \sim <\mathcal{M}_a^2> = 0.3$. It is smaller than our $f_\mathrm{crit}^{\dw}$ in particular for smaller $r_\mathrm{sh}$, which may indicate that our conclusion needs to be reconsidered. It should be noted, however, that the basic pictures of two criteria are qualitatively different. As we have explained in Sec.~\ref{sec:intro}, what we consider in this paper is an impulsive effect of progenitor asymmetry to shock revival. Indeed, the semi-dynamical approach on \citetalias{CritFluc} measures how large impulsive change of post-shock pressure is required to trigger shock expansion. On the other hand, the critical Mach number measures the role of progenitor asymmetry with quasi-steady contribution. This means that the picture of critical Mach number is relevant to the other role of progenitor asymmetry in which progenitor fluctuations couple with fluid instabilities in the post-shock region and then enhance the turbulent pressure (see also Sec~\ref{sec:intro}). The longer contribution of progenitor asymmetry results in reducing the required amplitude of fluctuations, which would be the main reason why the critical Mach number predicts the smaller pressure fluctuation than the critical fluctuation.

%to understand the difference between our creterion and the critical Mach number.

%, which results inrequired amplitude of fluctuations
% The long term contribution of progenitor asymmetry 
% The quasi-steady contribution to energize the post-shock flows results in smaller required amplitude of fluctuation than the critical fluctuation (see also \citet{Couch3} for the same argument.)

% and this study only focuses on the impulsive role of the asymmetric fluctuations.

Our current study is a first step to measure the importance of progenitor asymmetry by using a semi-analytic approach. We have in mind to extend our method to include quasi-steady contributions with multi-dimensional effects. The idea of critical Mach number $<\mathcal{M}_a^2>$ and also other diagnostics as the ante sonic conditions \citep{antesonic} and the integral-condition \citep{MM} will be important guidelines for the improvement. This work is currently underway and will be presented in the forthcoming paper.

\section{Summary and discussion}\label{sec.sum}
In this paper, we assess the importance of progenitor asymmetries to shock revival, in particular, focus on the impulsive role of progenitor asymmetry to trigger the shock revival. We test the scenario by employing a newly developed semi-analytical method. We first improve the classical quasi-steady model for the post-bounce phase of CCSNe by including progenitor dependence into the characteristic quantities as neutrino luminosity, mass accretion rate and PNS mass. We also calibrate free parameters in our model by comparing our model to the results of numerical simulations. These efforts allows us to make the time evolution of shock radius ($r_\mathrm{sh}$) and mass of PNS ($M_\mathrm{PNS}$) be reasonably consistent with those in numerical simulations. The two outputs, $r_\mathrm{sh}$ and $M_\mathrm{PNS}$, from the quasi-steady model are used to compute the critical fluctuation ($f_\mathrm{crit}^{\dw}$) which is the required amplitude of pressure fluctuations at the post-shock location for shock revival. After coverting $f_\mathrm{crit}^{\dw}$ to the corresponding fluctuation at the upstream ($f_\mathrm{crit}^\up$), we connect $f_\mathrm{crit}^\up$ to the progenitor asymmetry before the onset of collapse ($f_\mathrm{crit}^\mathrm{Si/O}$) by taking into account the growth of fluctuation during infall.

 We apply the semi-analytical model to two representative CCSNe progenitors models for $15$ and $27 M_\odot$ from realistic stellar evolution in \citet{WHW02} and six parametrically generalized progenitors in \citetalias{YY}. We find that the required progenitor asymmetry at the presupernova stage is $0.3~\lesssim~f_\mathrm{crit}^\mathrm{Si/O}~\lesssim~0.4$ for all the progenitors, which is roughly three times larger than the prediction by current stellar evolution models. We thus conclude that progenitor asymmetries can not trigger the shock revival by the impulsive way in the context of neutrino heating mechanism. It should be noted that there is an important caution in our conclusion that if the nuclear burning accelerates the growth of asymmetries during infall, the progenitor asymmetry could be a primary factor for the shock revival. This issue should be addressed in the forthcoming paper. 

Even though the progenitor fluctuations do not play a primary role to trigger a shock revival, their contribution to the shock revival is no doubt. As witnessed in recent detailed numerical simulations, the progenitor fluctuations promote fluid instabilities and turbulences in the post-shock flows. Although current semi-analytical models including ours are still unmatured, further improvements of the quasi-steady model and developments of better creterion for shock revival would allow us to assess the impact of progenitor asymmetries for various type of progenitors systematically. Efforts to develop better phenomenological model are as important as improving first-principle approach of CCSNe. Both developments are complement each other and lead us to comprehensive understanding of the explosion mechanism of CCSNe.

%Our approach would be applicalble if our model appropriately treat multi-dimensional effects

% which quasi-steadily push the shock wave outward

% can be frequently seen in more 

% Indeed, we found that the fluctuations in inner mass shells would disturb shock dynamics more than those in outer shells.

% we also concluded that one should give attention not only to the amplitude of fluctuations but also to their {\it locations} at the pre-supernova stage. This is another important outcome to understand the link between the stellar structures at the pre-supernova stage and the explosion mechanism.

%As discussed in Sec.~\ref{sec.limitation}, our model could be improved although the conclusions will not change. It is also possible to extend our spherically symmetric quasi-steady model to include multi-D effects as done by \cite{MM11}. Improved models would help us to understand the role of progenitor fluctuations in shock revival in more detail. We will address these issues in forthcoming papers.

\section*{Acknowledgments}
%We thank Wakana Iwakami for providing numerical routines related to EoS and weak interactions. 
We are grateful to Adam Burrows, Sherwood Richers, Jonathan Squire and Wakana Iwakami for valuable comments on this paper. We also appreciate the anonymous referee for his/her comments, which crucially helped us to improve this paper. This work is partially supported by a Research Fellowship for Young Scientists from the Japan Society for the Promotion of Science (JSPS). Hiroki Nagakura was supported in part by JSPS Postdoctoral Fellowships for Research Abroad No.~27-348 and he was partially supported at Caltech through NSF award No.~TCAN AST-1333520 and DOE SciDAC4 Grant DE-SC0018297 (subaward 00009650).



%\bibliographystyle{mnras}

%% \begin{thebibliography}{99}
%% \bibitem[\protect\citeauthoryear{Author}{2012}]{Author2012}
%% Author A.~N., 2013, Journal of Improbable Astronomy, 1, 1
%% \bibitem[\protect\citeauthoryear{Others}{2013}]{Others2013}
%% Others S., 2012, Journal of Interesting Stuff, 17, 198
%% \end{thebibliography}

%Alternatively you could enter them by hand, like this:
%This method is tedious and prone to error if you have lots of references

%%%%%%%%%%%%%%%%%%%%%%%%%%%%%%%%%%%%%%%%%%%%%%%%%%

%%%%%%%%%%%%%%%%% APPENDICES %%%%%%%%%%%%%%%%%%%%%

\appendix
\section{Calibration of free parameters in the quasi-steady model} \label{sec.calb}

\begin{figure*}
\begin{tabular}{cc}
\begin{minipage}{0.45\hsize}
%\begin{center}
%\includegraphics[bb = 0 0 461 346, width = \columnwidth]{tpb-rsh_comp.png}
%\end{center}
\includegraphics[width = \columnwidth]{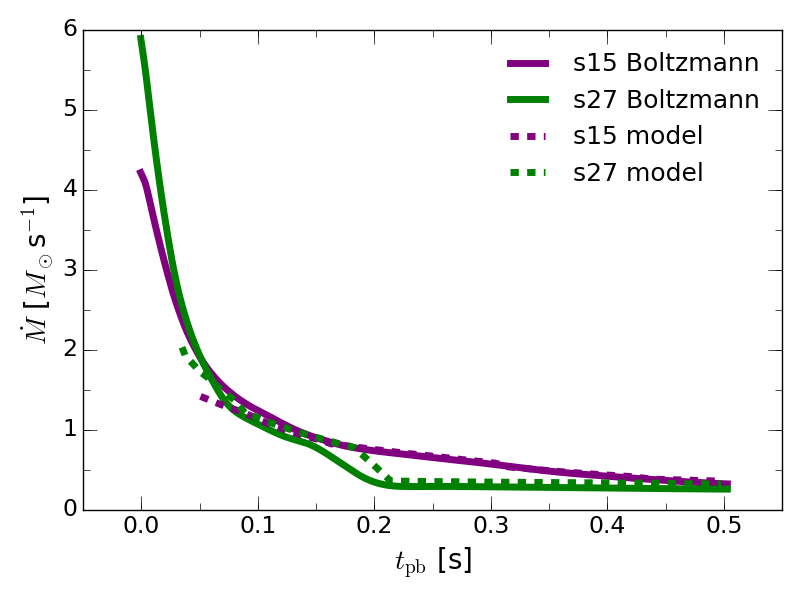}
\end{minipage} &
\begin{minipage}{0.45\hsize}
%\begin{center}
%\includegraphics[bb = 0 0 461 346, width = \columnwidth]{tpb-Mdot_comp.png}
%\end{center}
\includegraphics[width = \columnwidth]{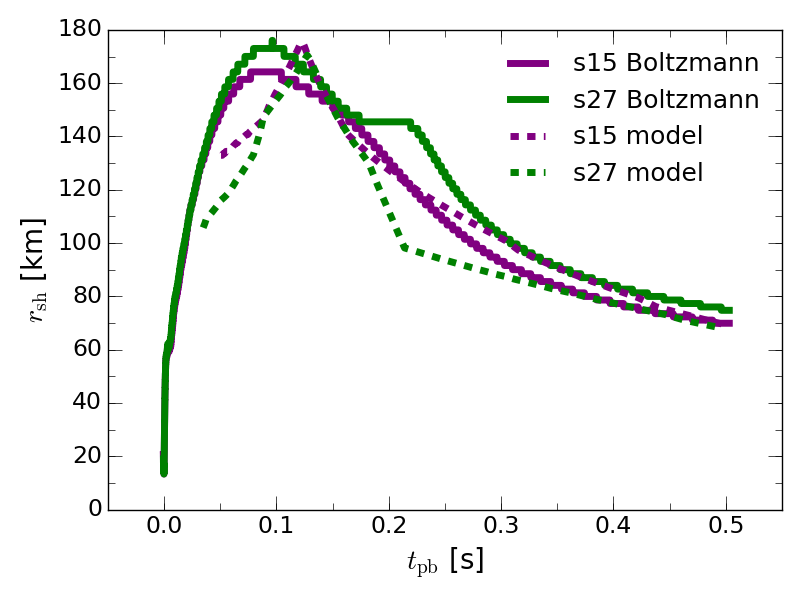}
\end{minipage}
\end{tabular}
\caption{The left and right panels show the time evolutions of mass accretion rate and shock radius, respectively. The solid lines show the results of 1D numerical simulations while the dashed ones show the result of our semi-analytical model with the fiducial set of parameters.}
\label{fig.comp}
\end{figure*}

The quasi-steady model with light bulb approximation is very useful to analyze the dynamics of post-bounce phase of CCSNe qualitatively. Indeed, it gave us an idea of critical neutrino luminosity which is frequently used to diagnose the closeness of explosions in the results of numerical simulations. Quantitatively speaking, however, the classical quasi-steady model is not enough accurate to reproduce the time evolution of CCSNe due to many simplifications. On the other hand, we need one as accurate as possible for the purpose of our study. This motivates us to improve the model. As shown below, the quasi-steady model is well improved by introducing several free parameters with being calibrated by the result of numerical simulations. 
%In this appendix, we show the calibration in detail.

Before describing the procedure of calibration in detail, we briefly explain the numerical setup and input physics of the reference simulations. The simulations were performed by the most up-to-date version of CCSN code in a Japanese group, while one of the authors in this paper is a main developer for this scheme \citep{Nagakura14,2017ApJS..229...42N,Nagakura18a}. Although the code is capable of solving multi-dimensional neutrino radiation hydrodynamic equations with full Boltzmann neutrino transport, we only use the result of spherically symmetric simulations to calibrate our semi-analytic model for the purpose of this paper. Note that the input physics in this code have been recently improved substantially, for instances, nuclear weak interactions as electron captures of heavy and light nuclei are consistently treated by multi-nuclear EoS. The EoS also combines with the variational method with the AV 18 (for two-body) and UIX (for three-body) nuclear potentials \citep{Togashi14,Togashi17,Furusawa17} for uniform matter. We use the results of simulations for s15 and s27 progenitors. We refer \citet{Nagakura18b} for more detail of the numerical simulations.

% The simulations were performed up to $t_{\rm pb} = 500$ms for these progenitors.

% The EOS for the supranuclear density is employeed
%  are summarized in our recent paper \citep{}.
% applicable for multi-dimensional CCSNe simulations 

% the results of numerical simulations which we employ 

The free parameters which we need to calibrate are, $\alpha$ in Eq.~(\ref{tff}), $\eta$ in Eq.~(\ref{Lacc}), $L_{\nu,\mathrm{diff}}^\mathrm{ref}$ and $\dot{L}$ in Eq.~(\ref{Ldiff}). The procedure of the calibration is as follows. We first search for a set of best fit parameters for s15 and s27 progenitor models independently. Note that the best fit parameters turn out to be almost identical, so we define the fiducial parameter by taking an average of each parameter. 

We at first search for $\alpha$ so as to reproduce the time evolution of $\dot{M}$ in numerical simulations. The $\alpha$ relates with $\dot{M}$ through Eq.(\ref{Mdot2}) in our model. The left panel of Fig.~\ref{fig.comp} compares the result of our fiducial parameter ($\alpha=1.5$) with those from numerical simulations. As shown in this panel, our model reasonably reproduces the time evolution of $\dot{M}$ in numerical simulations. Note that the time evolution of $M_\mathrm{PNS}$ in our model is also consistent with numerical simulations, since $\dot{M}$ dictates the increase of $M_\mathrm{PNS}$ with time.

Given $\alpha=1.5$, we then calibrate $L_{\nu,\mathrm{diff}}^\mathrm{ref}$ and $\dot{L}$ which are relevant to the diffusion component of neutrino luminosity. The decline rate of neutrino luminosity ($\dot{L}$) can be directly obtained from the time evolution of $L_{\nu}$ in numerical simulations and then $L_{\nu,\mathrm{diff}}^\mathrm{ref}$ is evaluated by extrapolating the decline rate to $t_{\rm pb} = 0$. Although the actual diffusion component would be different in particular at the early post-bounce phase, the error does not affect our model since the diffusion component is subdominant at that time.

% the neutrino luminosity in the early post-bounce phase is dominated by the accretion component, which means that 

Finally, we search for $\eta$ so as to reproduce the time evolution of shock radius. Note that, we do not attempt to reproduce the time evolution of neutrino luminosity in numerical simulations. We find that if we adopt the best fit $\eta$ to reproduce $L_{\nu}$ in numerical simulations, the obtained time evolution of shock radii are different from those of numerical simulation. This is simply because the light bulb approximation of neutrino transport is too simplified. Indeed, it assumes the thermal spectrum with zero chemical potential of neutrinos (which are not true in reality) and also ignores some dominant weak interactions as nucleon scatterings. Since it is not easy to improve the light bulb approximation and addressing this issue is beyond the scope of this paper, we decide to match $r_\mathrm{sh}$ from numerical simulations instead of $L_{\nu}$. Note that, the time evolution of $r_\mathrm{sh}$ is the most important quantity and the deviation of $L_{\nu}$ from realistic simulations is not a problem in this study.

 The right panel of Fig.~\ref{fig.comp} shows the result of time evolution of shock radius with best fit parameters, which are $\eta=0.69$, $L_{\nu,\mathrm{diff}}^\mathrm{ref}=1.7\times10^{52}$~erg~s${}^{-1}$, $\dot{L}=-3\times10^{52}$~erg~s${}^{-2}$. The panel also shows the results of numerical simulations for comparisons. We see some deviations between two results at $0.2{\rm s} \lesssim t_{\rm pb} \lesssim 0.3$s for s27 progenitor. This corresponds to the phase when the Si/O layers hit the shock wave. At this phase, the quasi-steady approximation would be invalid since the background changes more quickly than the time scale which the system settles down the quasi-steady state, i.e., more dynamical treatments are required to capture the trend quantitatively. Although this is an interesting issue, the improvement is the beyond the scope of this paper. Importantly, our semi-analytical model underestimates $r_\mathrm{sh}$ than those in numerical simulations, which means that the caveat does not change our conclusion (see Sec.~\ref{effrsh} for more details). Although there remains some issues as described above, the results displayed in Fig.~\ref{fig.comp} lead confident to our semi-analytical model.

\section{Parameter dependence} \label{appModel}

\begin{figure*}
\begin{tabular}{cc}
\begin{minipage}{0.42\hsize}
\includegraphics[width = \columnwidth]{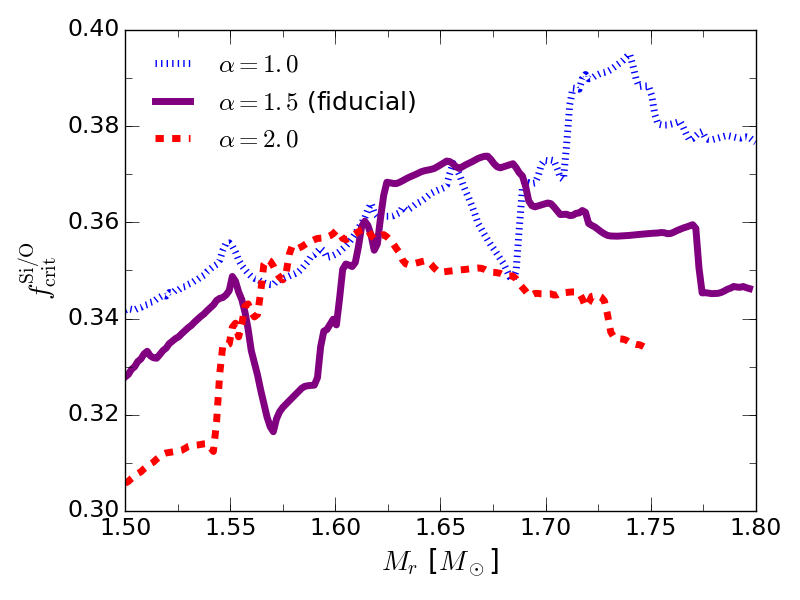}
\end{minipage} &
\begin{minipage}{0.42\hsize}
\includegraphics[width = \columnwidth]{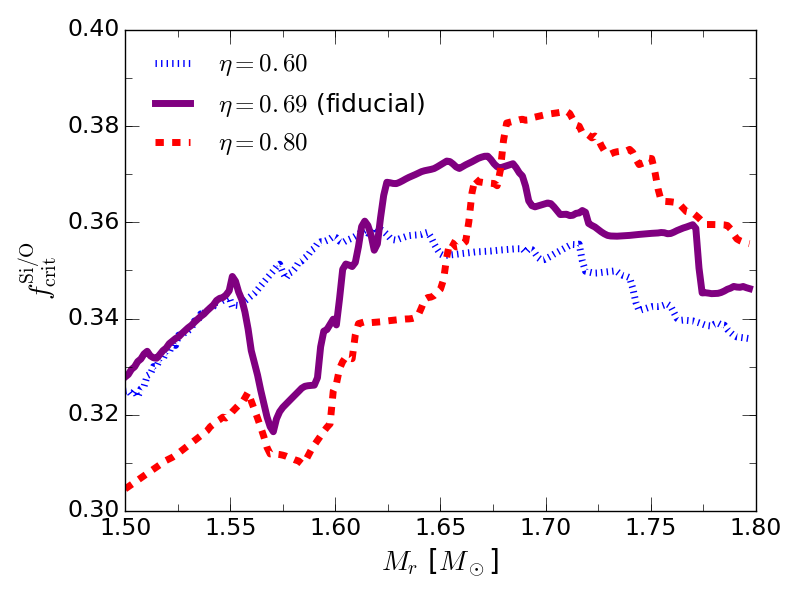}
\end{minipage} \\
\begin{minipage}{0.42\hsize}
\includegraphics[width = \columnwidth]{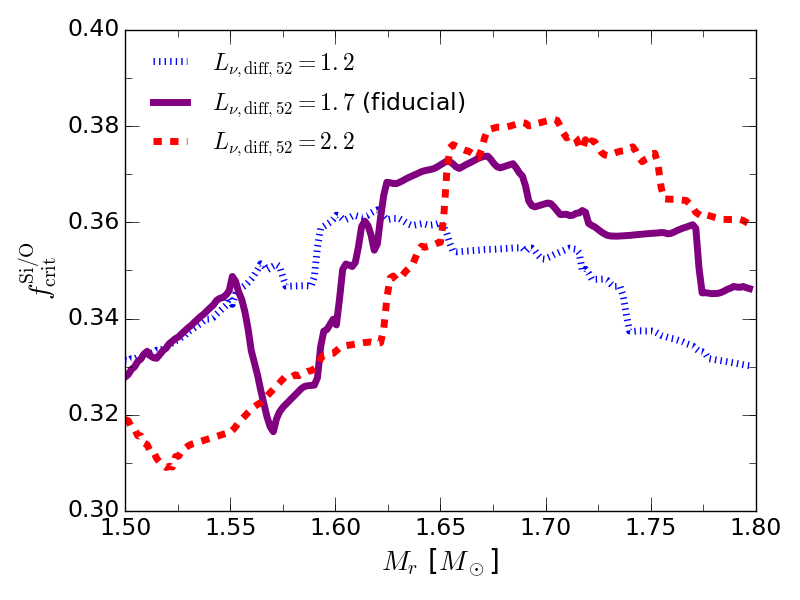}
\end{minipage} &
\begin{minipage}{0.42\hsize}
\includegraphics[width = \columnwidth]{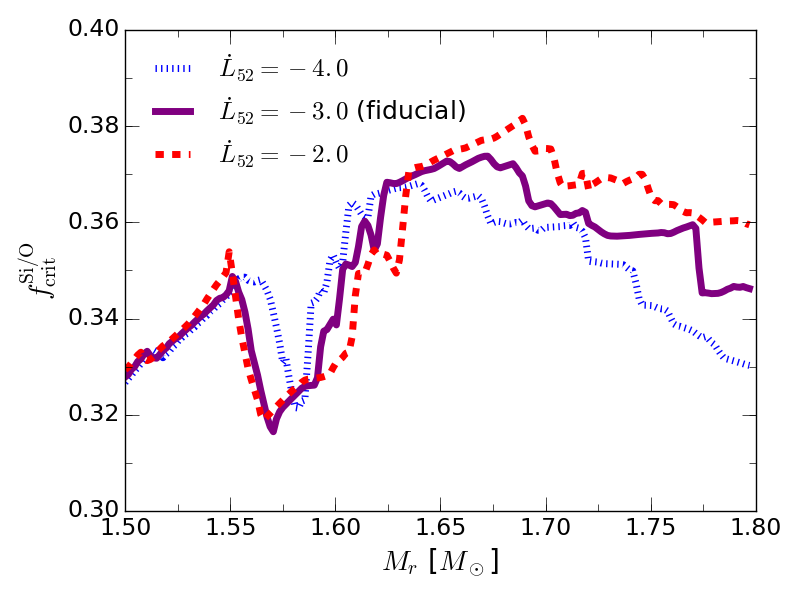}
\end{minipage}
\end{tabular}
\caption{Summary of the study of parameter dependence. Top left and right are for $\alpha$ and $\eta$, respectively. Bottom left and right are for $L_{\nu,\mathrm{diff}}^\mathrm{ref}$ and $\dot{L}$, respectively. We display $f_\mathrm{crit}^\mathrm{Si/O}$ as a function of $M_r$.}
\label{fig:paradepe}
\end{figure*}

%Although we calibrate the free parameters in our model so as to reproduce the results of numerical simulations, it is meaningful to study

We examine the parameter dependences in our moel. We study each parameter dependence by fixing others to fiducial values. We summarize the result in Fig.~\ref{fig:paradepe} which shows $f_\mathrm{crit}^\mathrm{Si/O}$ as a function of $M_r$. 
%Below, we discuss the qualitative trend of each parameter dependence.
% Fig.~\ref{fig:paradepe} summarizes the dependence of 

We first consider $\alpha$ dependence of $f_\mathrm{crit}^\mathrm{Si/O}$ (the top left panel in Fig.~\ref{fig:paradepe}). The $\alpha$ is a primary parameter which determines the overall magnitude of $t_{\rm pb}$ as well as $\dot{M}$. Larger $\alpha$ leads to a slower evolution of the system (since $t_{\rm pb}$ tends to be larger) with the smaller mass accretion rate. Roughly speaking, larger $\alpha$ leads smaller $f_\mathrm{crit}^\mathrm{Si/O}$, which is clearly seen in the mass shell of $M_r>1.7M_\odot$. This trend can be understood as follows. As mentioned already, larger $\alpha$ prolongs the time scale of post-bounce phase, in other words, the same mass shell hits the shock wave at later. As a result, the diffusion component of neutrino luminosity becomes smaller (see also Eq.~(\ref{Ldiff})). The reduction of $L_{\nu}$ results in decreasing $r_\mathrm{sh}$. Since the smaller $r_\mathrm{sh}$ prolongs the supersonic accretion phase for infalling matter, the progenitor fluctuation is more amplified during infall. As a result, $f_\mathrm{crit}^\mathrm{Si/O}$ becomes smaller\footnote{Note that the amplification during infall dominates the critical fluctuation for $r_\mathrm{sh} \lesssim 100$km. See Sec.~\ref{effrsh} in more detail.}. Hence, the calibration of $\alpha$ turns out to be very important to consider the impact of progenitor asymmetry in particular for the late phase (for the larger mass shell). The impact to $f_\mathrm{crit}^\mathrm{Si/O}$ can be estimated as $\sim 20\%$ in $1<\alpha<2$.

The $\eta$ dependence is displayed in the top right panel in Fig.~\ref{fig:paradepe}. It determines the conversion efficiency from accretion energy to neutrino emission (see Eq.~(\ref{Lacc})). Although $r_\mathrm{sh}$ becomes monotonically larger with larger $\eta$, $f_\mathrm{crit}^\mathrm{Si/O}$ depends on $\eta$ in a more complex way. This is mainly due to the fact that, as discussed in Sec.~\ref{effrsh}, there is a competition between two effects which the amplification during infall and the critical fluctuation. In the early phase, the larger $r_\mathrm{sh}$ reduces $f_\mathrm{crit}^\mathrm{Si/O}$ since the decrease of critical fluctuation dominates the suppress of amplification factor during infall. On the contrary, the dominance is reversed in the late phase. For this reason, the larger $r_\mathrm{sh}$ in larger $\eta$ models increases $f_\mathrm{crit}^\mathrm{Si/O}$.

% become larger than the fiducial model.
% It determine $f_\mathrm{crit}^\mathrm{Si/O}$.

 Note that, for $L_{\nu,\mathrm{diff}}^\mathrm{ref}$ dependence (see the bottom and left panel in Fig.~\ref{fig:paradepe})), the dependence is almost the same as $\eta$ dependence since $r_\mathrm{sh}$ increases with monotonically with increase of $L_{\nu,\mathrm{diff}}^\mathrm{ref}$. On the other hand, for $\dot{L}$ dependence, the systematic trend can be seen for $M_r>1.65M_\odot$. The smaller $\dot{L}$ leads higher $L_{\nu}$ and then larger $r_\mathrm{sh}$. The difference is more prominant in the late phase. Again, the increase of $r_\mathrm{sh}$ reduces $f_\mathrm{crit}^\mathrm{Si/O}$ at the late phase due to the same reason as mentioned in other parameter dependences.

\bsp	% typesetting comment
\label{lastpage}
\end{document}